\documentclass[iop]{emulateapj}
\usepackage{multirow}
\usepackage[english]{babel}
\usepackage{natbib}
\usepackage[usenames,dvipsnames]{color}

\slugcomment{ApJ, in press}
\shorttitle{ETGs at z $\sim$ 1.3. IV. Scaling relations in different environments}
\shortauthors{Raichoor et al.}

\usepackage{hyperref}

\begin{document}
\title{Early-type galaxies at z $\sim$ 1.3. IV. Scaling relations in different environments}

\author{
A. Raichoor\altaffilmark{1,17}, 
S. Mei\altaffilmark{1,2,3},
S.A.Stanford\altaffilmark{4,5,18},
B.P. Holden\altaffilmark{6},
F. Nakata\altaffilmark{7},
P. Rosati\altaffilmark{8},
F. Shankar\altaffilmark{9},
M. Tanaka\altaffilmark{10},
H. Ford\altaffilmark{11},
M. Huertas-Company\altaffilmark{1,2},
G. Illingworth\altaffilmark{6},
T. Kodama\altaffilmark{7,12},
M. Postman\altaffilmark{13},
A. Rettura\altaffilmark{4,11,14},
J.P. Blakeslee\altaffilmark{15},
R. Demarco\altaffilmark{16},
M.J. Jee\altaffilmark{4},
R.L. White\altaffilmark{13}}

\altaffiltext{1}{GEPI, Observatoire de Paris, Section de Meudon, 5 Place J. Janssen, 92190 Meudon Cedex, France} 
\altaffiltext{2}{Universit\'{e} Paris Denis Diderot, 75205 Paris Cedex 13, France}
\altaffiltext{3}{California Institute of Technology, Pasadena CA 91125, USA}
\altaffiltext{4}{Department of Physics, University of California, 1 Shields Avenue, Davis, CA 95616, USA}
\altaffiltext{5}{Institute of Geophysics and Planetary Physics, Lawrence Livermore National Laboratory, Livermore, CA 94551, USA}
\altaffiltext{6}{UCO/Lick Observatories, University of California, Santa Cruz 95065, USA}
\altaffiltext{7}{Subaru Telescope, National Astronomical Observatory of Japan, 650 North A{'}ohoku Place, Hilo, HI 96720, USA}
\altaffiltext{8}{European South Observatory, Karl-Schwarzschild -Str. 2, D-85748, Garching, Germany}
\altaffiltext{9}{Max-Planck-Instit\"{u}t f\"{u}r Astrophysik, Karl-Schwarzschild-Str. 1, D-85748, Garching, Germany}
\altaffiltext{10}{Institute for the Physics and Mathematics of the Universe, The University of Tokyo, 5-1-5 Kashiwanoha, Kashiwa-shi, Chiba 277-8583, Japan}
\altaffiltext{11}{Department of Physics and Astronomy, Johns Hopkins University, Baltimore, MD 21218, USA}
\altaffiltext{12}{National Astronomical Observatory of Japan, Mitaka, Tokyo 181-8588, Japan}
\altaffiltext{13}{Space Telescope Science Institute, 3700 San Martin Drive, Baltimore, MD 21218, USA}
\altaffiltext{14}{Department of Physics and Astronomy, University of California, Riverside, CA 92521, USA}
\altaffiltext{15}{Herzberg Institute of Astrophysics, National Research Council of Canada, Victoria, BC V9E 2E7, Canada}
\altaffiltext{16}{Department of Astronomy, Universidad de Concepci\'{o}n, Casilla 160-C, Concepci\'{o}n, Chile}
\altaffiltext{17}{\textit{Current address}: Osservatorio Astronomico di Brera, via Brera 28, 20121 Milan, Italy; e-mail: anand.raichoor@brera.inaf.it} 
\altaffiltext{18}{Visiting Astronomer, Kitt Peak National Observatory, National Optical Astronomy Observatories, which is operated by the Association of Universities for Research in Astronomy, Inc. (AURA) under cooperative agreement with the National Science Foundation.}

\begin{abstract}

  We present the Kormendy and mass-size relations for early-type galaxies (ETGs) as a function of environment at $z \sim1.3$. Our sample includes 76 visually classified ETGs with masses $10^{10} < M/M_{\sun} < 10^{11.5}$, selected in the Lynx supercluster and in the GOODS/CDF-S field, 31 ETGs in clusters, 18 in groups and 27 in the field, all with multi-wavelength photometry and \textit{HST}/ACS observations.  The Kormendy relation, in place at $z \sim1.3$, does not depend on the environment. 
The mass-size relation reveals that ETGs overall appear to be more compact in denser environments: cluster ETGs have sizes on average around 30-50\% smaller than those of the local universe, and a distribution with a  smaller scatter, whereas field ETGs show a mass-size relation with a similar distribution than the local one. Our results imply that (1) the mass-size relation in the field did not evolve overall from $z \sim 1.3$ to present; this is  interesting and in contrast to the trend found at higher masses from previous works; (2) in denser environments, either ETGs have increased their size by 30-50\%, on average, and spread their distributions, or more ETGs have been formed within the dense environment from not ETG progenitors or larger galaxies have been accreted to a pristine compact population to reproduce the mass-size relation observed in the local Universe.
Our results are driven by galaxies with masses $M \lesssim 2 \times 10^{11} M_{\sun}$ and those with masses $M \sim 10^{11} M_{\sun}$ follow the same trends that the entire sample.
Following Valentinuzzi et al. definition of superdense ETGs, around 35-45\%  of our cluster sample is made of superdense ETGs.

\end{abstract}
\keywords{galaxies: clusters: individual (RX J0849+4452, RX J0848+4453)  -- galaxies: elliptical and lenticular -- galaxies: evolution -- galaxies: formation -- galaxies: high-redshift -- galaxies: fundamental parameters}


\section{Introduction}

In recent years, studies have unveiled the existence at $z \sim$ 1-2 of a population of massive spheroidal galaxies with small size, hence called compact \citep[e.g.,][and also references therein]{daddi05,trujillo06,trujillo07,buitrago08,cimatti08,van-der-wel08,van-dokkum08,damjanov09,saracco09,newman10,rettura10,saracco10,strazzullo10}.  When comparing those high redshift galaxies with local ones of similar mass, it appears that their sizes are smaller by a factor of $\sim$2-3 and up to 5 \citep{van-dokkum08}.  The general view is that the compactness increases with redshift, mass and the level of quiescence \citep[e.g.,][]{trujillo07,franx08,williams10}.  
Despite potential selection biases affecting the comparison of high vs low redshift samples (see hereafter), which might affect conclusions on the evolution in size, the existence of a significant number of compact galaxies at high redshift is firmly established.
\footnote{We underline that not all high redshift ETGs are compact  \citep[e.g.,][]{mcgrath08,saracco09,mancini10,onodera10, saracco11}}.
 
The presence of compact ETGs in the local Universe is still debatable. Apparent disagreements may come from the different definitions for a compact galaxy (i.e. the different mass and size criteria chosen to define a galaxy as compact).  For example, on the one hand, the analysis of Sloan Digital Sky Survey \citep[SDSS; ][]{york00} samples reveals that a negligible fraction of galaxies are compact \citep{trujillo09}, even when taking into account the possible incompleteness due to the SDSS spectroscopic target selection algorithm \citep{taylor10}.  On the other hand, \citet{valentinuzzi10} studied ETGs in local clusters and found that a significant fraction of their sample is made of compact objects.  When compared to high redshift samples \citep{saracco09}, the number density of compact galaxies at $1 < z < 2$ is consistent with that found in this last work and consistent with a lack of evolution in size \citep[see also][]{shankar10a,bernardi10}.

The formation of compact galaxies might be a consequence of mergers of gas-rich subunits at high redshift \citep[e.g.,][]{khochfar06a,hopkins09b,wuyts10} and/or cold flows \citep[e.g.,][]{bournaud11}, resulting in an intense starbust and compact quiescent remnant due to highly dissipative processes. This is in agreement with observations showing that the gas fraction of star-forming galaxies increases with redshift \citep{hopkins10}.  Sub-millimeter galaxies have been suggested as promising candidates for compact galaxies precursors \citep{granato06, cimatti08}. The picture concerning the subsequent evolution of compact galaxies down to $z=0$ is more difficult to draw.  
The comparison of high redshift to local samples may be affected by two selection biases: age selection bias against young galaxies in high redshift samples \citep[e.g.][]{saglia10,valentinuzzi10} and progenitor bias due to morphological evolution \citep[e.g.,][]{van-dokkum01a,kaviraj09,valentinuzzi10a}. Within this context, it is still unclear which part of the galaxy population went through evolution and which mechanism contributed to it.
If the compact galaxy population requires evolution, one efficient process may be minor dry mergers \citep{naab09,shankar11}: through the accretion of gas-poor satellites, a compact galaxy will increase significantly its size with a limited increase of its mass and no star formation.  In this scenario, the accreted material will extend the outer parts of the compact galaxy, leaving its core unchanged.
This is in remarkable agreement with observations: local elliptical galaxies have in their core regions surface stellar density profiles similar to those of high-redshift compact galaxies \citep[e.g.,][]{bezanson09,hopkins09,van-dokkum10}.
Another proposed scenario for size evolution of compact spheroids is expansion consequent to substantial mass losses due to, e.g., stellar winds and/or quasar feedback \citep{fan08}.

To go deeper in understanding these mechanisms, it is useful to study the mass-size relation as a function of environment. 
Until now, few studies have covered the full range of environment when studying the mass-size relation at $z > 1$.
In the local universe, \citet{maltby10} have found that the mass-size relation does not depend on the environment for ETGs.
Most of the current $z > 1$ studies though rely on field samples, except for \citet{rettura10} and \citet{strazzullo10}, who studied clusters at $z \sim$ 1.2-1.4.
Only \citet{rettura10} compared field and cluster ETGs at $z \sim 1.2$ and find that galaxies from different environments lie on the same relations. 

 In \citet[][R11 herafter]{raichoor11}, we presented a unique homogeneous sample of ETGs probing cluster, group and field environments at $z \sim 1.3$.  Our study relies on high-quality multi-wavelength data covering the Lynx supercluster \citep{stanford97,rosati99,nakata05,mei06,mei11,rettura11}, a structure at $z = 1.26$ made of two clusters and at least three groups.  
From our spectroscopic runs on the groups, we obtained average spectroscopic redshift $z=1.262 \pm 0.007$ (Group~1; from 9 members), $z=1.260 \pm 0.006$ (Group~2; from 7 members), $z=1.263 \pm 0.005$ (Group~3; from 9 members) \citep{mei11}.
Group X-ray emission gives masses less or around $5 \times 10^{13} M_{\sun}$ \citep{mei11}.
Group 2 and 3 appear to be spatially separated (as from our Friend-of-Friend algorithm) from the two clusters, while Group 1 is spatially connected to the Lynx W cluster.
We consider it as a separate group though, because its center is at  $1.1 \times r_{200}$ from the center of the cluster and it extends to  $2 \times r_{200}$, with an area of very low density between 0.5--1$\times r_{200}$.
It might be close to merging to Lynx W, or in the merger process.
For further details please refer to \citet{mei11}.
Our groups belong all to the Lynx supercluster, and are not isolated.
They do not show peculiar densities, or masses to differentiate them from isolated groups.
A more extended analysis of supercluster groups as compared to isolated groups at the same redshift would help us understanding if the properties of their galaxies might be different.
At the moment, we do not have elements to suggest it.

In this paper, we use the R11 sample to study the influence of environment on the structural parameters of ETGs at high redshift as a function of mass and environment.  The estimates of ETG sizes from \textit{HST}/ACS images combined with the photometry and the stellar population parameters determined in R11 allow us to build the two key relations to study structural parameters of ETGs: the \citet{kormendy77} relation (KR) and the mass-size relation (MSR).  

The plan of this paper is as follows.  In $\S$2, we present the observations, the sample selection and the SED fitting method used to estimate ages and masses.  In $\S$3, we describe our estimation of the ETG structural parameters.  In $\S$4, we study the Kormendy relation and in $\S$5 the mass-size relation.  We then present our conclusions in $\S$6.
	
We adopt a standard cosmology with $H_0 = 70$ km s$^{-1}$ Mpc$^{-1}$, $\Omega_m = 0.30$ and $\Omega_\Lambda = 0.70$.  All magnitudes are in the AB system.
Unless otherwise stated, all stellar masses are computed with a \citet{salpeter55} Initial Mass Function (IMF).  We choose as our rest-frame reference the Coma cluster ($z_0 = 0.023$).


\section{Observations, sample selection, photometry and SED fitting \label{sec:osps}}
This work relies on optical and infrared (0.6-4.5 $\mu$m) images of the Lynx supercluster and of the Great Observatories Origins Deep Survey \citep[GOODS;][]{giavalisco04} observations of the Chandra Deep Field South \bibpunct[; ]{(}{)}{;}{a}{}{;} \citep[CDF-S;][Dickinson et al., in preparation]{giavalisco04,nonino09,retzlaff10}.
\bibpunct{(}{)}{;}{a}{}{,}
The observations, the sample selection, the photometry and the age and stellar mass estimation are presented in R11 and we briefly summarize them here; please refer to R11 for more details. 
The images cover seven bandpasses: $R$ (Keck/LRIS for the Lynx clusters, Palomar/COSMIC for the Lynx groups, VLT/VIMOS for the CDF-S), \textit{HST}/ACS F775W and F850LP -- hereafter $i_{775}$ and $z_{850}$, $J$/$K_s$ (KPNO/FLAMINGOS for the Lynx clusters and groups, VLT/ISAAC for the CDF-S), \textit{Spitzer}/IRAC ch1 and ch2 -- hereafter $[$3.6$\mu$m$] $ and $[$4.5$\mu$m$]$.
The sample of R11 consists of 79 ETGs (31 in the Lynx clusters, 21 in the Lynx groups and 27 in the CDF-S) selected in redshift ($0.92 \le z_{phot} \le 1.36$ for the Lynx ETGs and $1.1 \le z_{spec} \le 1.4$ with $\langle z_{spec} \rangle = 1.239 \pm 0.082$ for the CDF-S), in magnitude (21 $\le z_{850}$ (AB) $\le 24$) and in morphology (E/S0 types based on visual inspection of $z_{850}$-band of \textit{HST}/ACS images according to \citet{postman05} and \citet{mei11} classification). 
We verified that the magnitude cut $z_{850} \ge 21$ does not exclude any galaxy satisfying the $z_{phot}$ or $z_{spec}$ selection criteria; thus we can relax the magnitude cut to $z_{850}$ (AB) $\le 24$ without affecting the sample.  ETGs belonging to the Lynx clusters and groups are identified in \citet{mei11} by a Friend-Of-Friend algorithm \citetext{FOF, \citealp{geller83}; see also \citealp{postman05}} with a linking scale corresponding to a local distance of $0.54$~Mpc, normalized to $z = 1.26$ and to our magnitude range \citep{postman05,mei11}.
We also verified that the selected CDF-S ETGs are field ETGs, i.e. that they do not belong to already identified structures (see R11).

At $z_{850} = 24$ mag, Lynx samples are complete and our CDF-S sample is more than $70\%$ complete (see R11).
The Lynx cluster, group, and CDF-S field samples have similar spectral coverage and are almost complete at $z_{850} = 24$ mag, thus providing a homogeneous and consistent sample.
Since the publication of R11, spectroscopic observations revealed that three ETGs from our Group 2 sample were outliers (ID = 939, 1791, 2519).
We thus remove those three ETGs from our sample, obtaining a final sample of 76 ETGs (31 in the Lynx clusters, 18 in the Lynx groups and 27 in the CDF-S).
The removal of those three outliers does not affect significantly any of the results presented in R11.
Our sample has spectroscopic redshifts for 20/31 ETGs in the clusters, 8/18 ETGs in the groups \citep{mei11} and 27/27 ETGs in the field.


We performed photometry in circular apertures with 1.5\arcsec~ radius and derive a  multi-wavelength photometric catalog with total magnitudes, determined using PSF growth curves. 
We estimated stellar masses and stellar population ages by fitting the SED with different stellar population models (\citet{bruzual03}, \citet{maraston05}, and an updated version [CB07] of \citet{bruzual03} that implements a new modeling of the TP-AGB phase). 
We hereafter refer to those models as BC03, MA05 and CB07, respectively. 
For SED fitting we used a \citet{salpeter55} IMF, solar metallicity, exponentially declining star-formation histories $\psi(t) \propto e^{-t/\tau}$ with a characteristic time $0.1 \le$ SFH $\tau$ (Gyr) $\le 5$, and no dust. 
Our stellar mass is the mass locked into stars, including stellar remnants\footnote{Using the nomenclature given by the authors: column 7 of *.4color files for BC03/CB07 models and "M$\wedge \ast$ total" for MA05 models} and our age is star-formation weighted age. 
A detailed discussion of different choices of parameters can be found in R11.


\section{Size estimation \label{sec:size}}

\subsection{Method}

In this section, we describe our methodology to derive the size of our ETGs.  Morphological parameters are usually estimated in the rest-frame $B$--band: we derive them from the \textit{HST}/ACS z$_{850}$ band image, the closest to the rest-frame $B$--band in our sample.  To fit the observed two-dimensional surface brightness distributions to a model, we use the software \textsc{Galfit} \citep[v3.0.2, ][]{peng02}, which has been shown to give reliable results \citep{haussler07}.  We assumed a \citet{sersic68} $r^{1/n}$ profile: \begin{equation} I(r) = I_e \times exp\{-b_n [(r/r_e)^{1/n}-1]\}, \end{equation} where $I(r)$ is the surface brightness at $r$, $I_e$ is the surface brightness at the effective radius $r_e$, which is the radius which encloses half of the emitted light.
In the fit, \textsc{Galfit} convolves the model with a provided PSF: 
our PSF stamp is built from real isolated unsaturated stars, by first normalizing them and then taking the median value for each pixel (the same as the one used in R11, see this paper for more details). It has been shown in the literature that considering different stars for the PSF leads to minor changes in the size estimate \citep[e.g.][]{trujillo07}.
\textsc{Galfit} outputs the semi-major axis $r_e$ of the projected elliptical isophote containing half of the total light and the axis ratio $b/a$.  Throughout this work, we use $R_e$ to denote the circularized effective radius defined by: \begin{equation}
  R_e = r_e \times \sqrt{b/a}.\\
\end{equation}

For each object, we create a square stamp from the ACS image centered on the galaxy.
According to our tests (using stamp size of $2.5 \times r_1$, $5 \times r_1$ and $10 \times r_1$), the fit is stable for a stamp size of $5 \times r_1$, where $r_1$ denotes the \citet{kron80} radius, as determined by SExtractor \citep{bertin96}.  We simultaneously fit the selected ETG along with any objects closer than 2.5\arcsec and use SExtractor segmentation maps to mask the other objects.  During the fit, we let as free parameters the position ($x$,$y$), the total magnitude $z_{850}$, the effective radius $r_e$, the axis ratio $b/a$, the S\'ersic index $n$ and the position angle $pa$.  While we use SExtractor outputs as initial guess for ($x$,$y$), $z_{850}$, $r_e$, $b/a$ and $pa$, we set the initial S\'ersic index $n$ to 2.5. 
As advised in the Galfit homepage, no boundary constraint on the S\'ersic index is provided during the fit, so that the minimization algorithm can run properly.
In order to reduce the number of free parameters and improve the quality of the fit, we fix the sky value.  For sky estimation, we create a larger stamp (20\arcsec$\times$20\arcsec) centered on the ETG, we mask the objects with ellipses (taking SExtractor's outputs and increasing the linear size by a factor 5) and take the median value of the remaining pixels.  This conservative approach for masking objects ensures that there is negligible residual light from the objects in the sky area, while keeping a large enough number of pixels.

For five ETGs of our sample ($\sim 6 \%$), our fits do not provide satisfactory results: either the output parameters are unphysical ($n > 10$, small $R_e$), or the residuals are unsatisfactory (two ETGs).
For those five ETGs, we consider the structural parameter estimates as non robust and we flag them in the figures in the paper.

\subsection{Reliability of the fit}

We test the robustness of our size estimation by applying the same fitting procedure to a set of simulated galaxies.
We generate 1,000 galaxies with randomly input magnitude ($21 \le z_{850,in} \le 24$), effective radius ($0.1\arcsec \le R_{e,in} \le 1.2\arcsec$), a S\'ersic index following a gaussian distribution $(\mu,\sigma) = (4,2)$ (with the constraint $n_{in} > 0.1$, to prevent from negative values), random position angle and axis ratios following a gaussian distribution $(\mu,\sigma) = (0.65,0.1)$.
The magnitude and axis ratio ranges are representative of our sample.
The range in S\'ersic index and effective radius is chosen according to the local distribution of ETGs \citep{caon93,blanton05,shankar10}.  

We then convolve the simulated galaxy with the PSF image and add Poissonian noise.
The simulated galaxy is eventually placed in a stamp extracted from the real \textit{HST}/ACS $z_{850}$ image, randomly chosen between ten positions devoided of sources, thus taking into account the background noise and all possible systematics inherent to the image.

\begin{figure*}
	\begin{tabular}{ccc}
	   \includegraphics[width=0.5\linewidth]{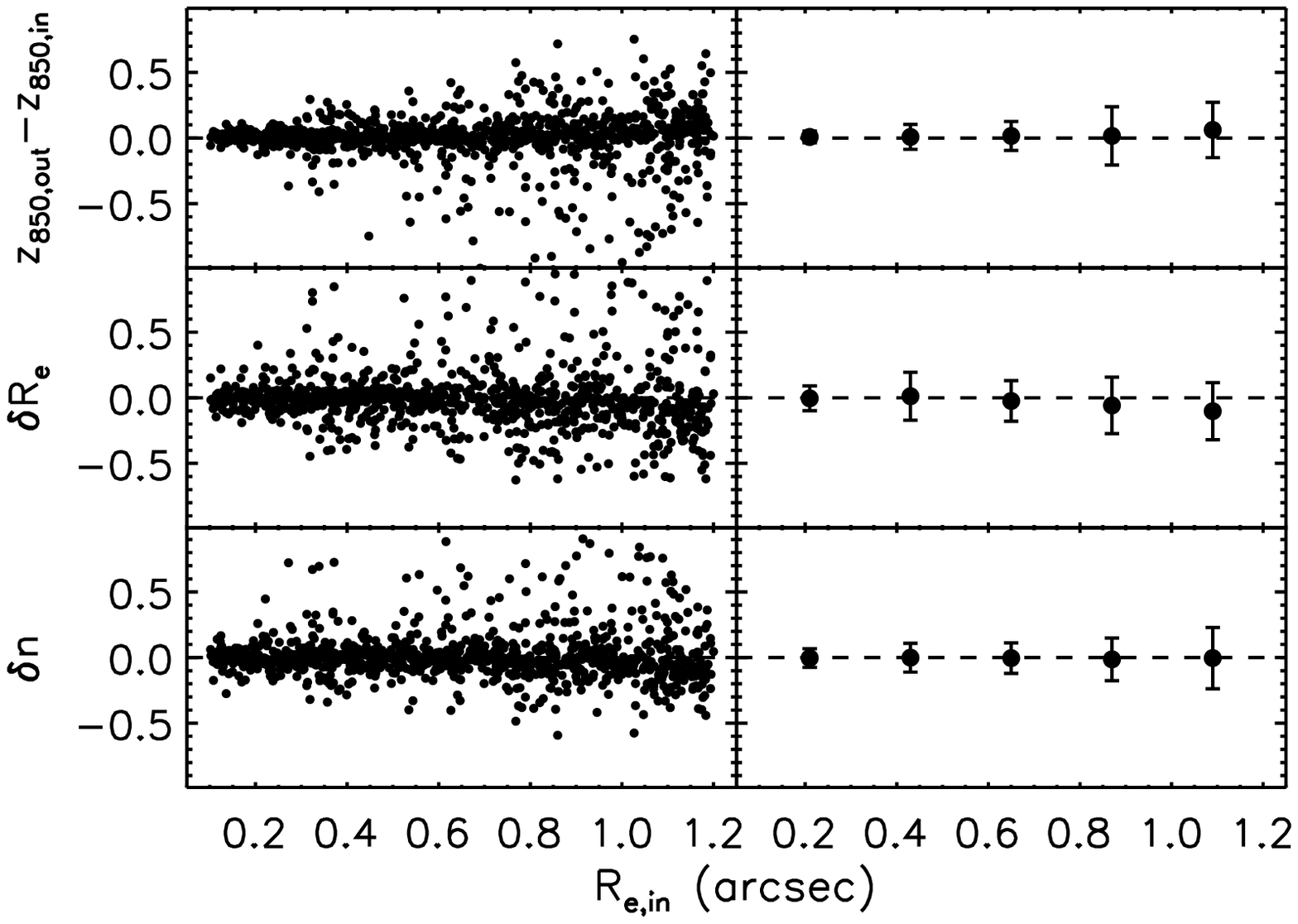} & 
	   \includegraphics[width=0.5\linewidth]{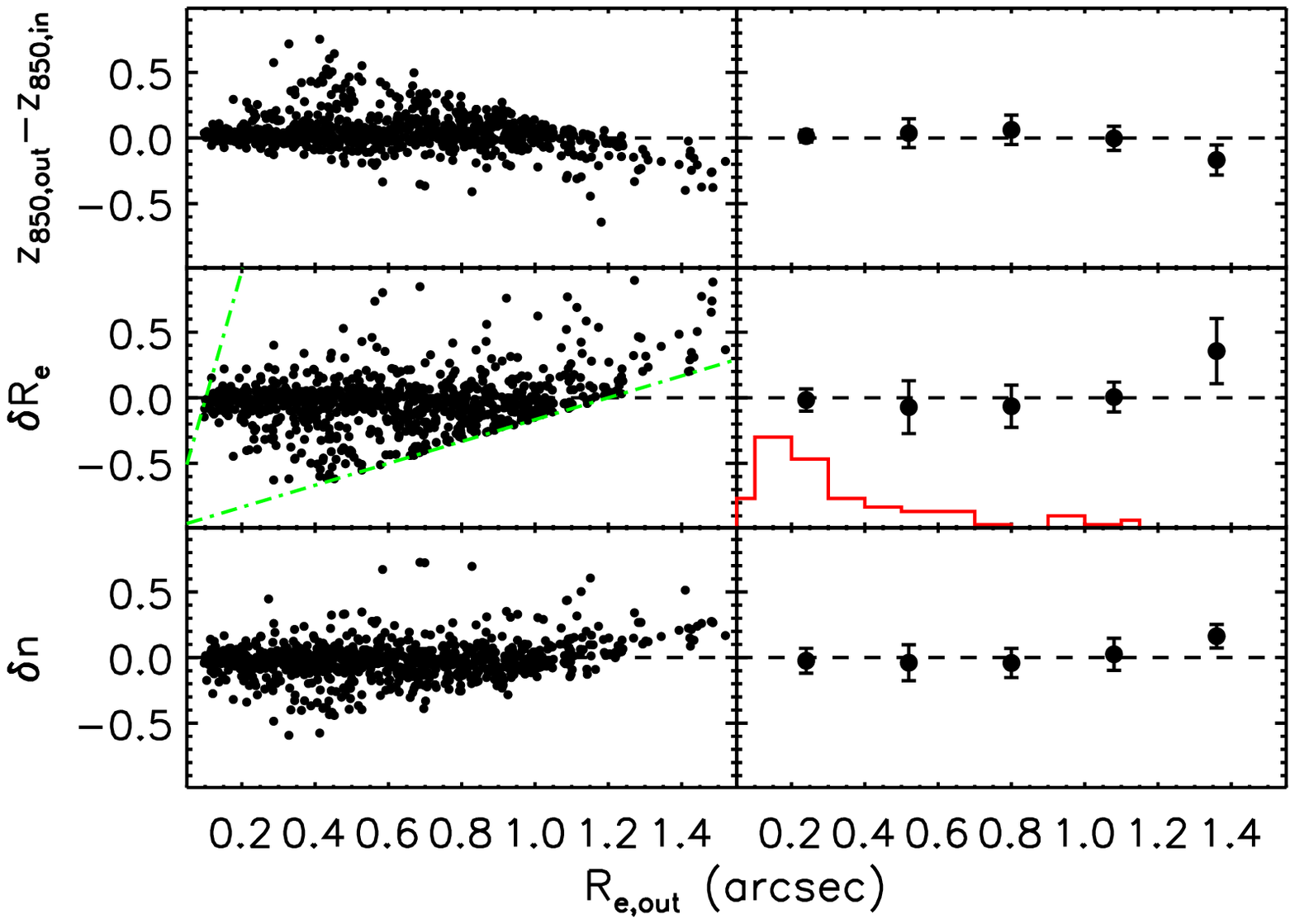} \\
	\end{tabular}
	\caption{Robustness of the fitting method:
comparison of the estimated and input parameters
(magnitude: $(z_{850,out}-z_{850,in})$,  
effective radius:  $\delta R_e = (R_{e,out}-R_{e,in})/R_{e,in}$, 
S\'ersic index:  $\delta n = (n_{out}-n_{in})/n_{in}$) 
versus the input effective radius $R_{e,in}$ (\textit{left panel}) and the measured effective radius $R_{e,out}$ (\textit{right panel}).
In the right part of each panel, we bin the values with x-axis value bins: the plots show the mean value and the scatter for each bin.
The red histogram in the right panel represents in arbitrary units the $R_e$ distribution for our real ETGs.
The green dashed lines delimit the possible $\delta R_e$ values due to the range of the simulation input values of $R_{e,in}$: because of the definition of $\delta R_e$ and of $0.1\arcsec \le R_{e,in} \le 1.2\arcsec$, a galaxy with $R_{e,out}$ will necessary have a corresponding value of $\delta R_e$ within [$R_{e,out}/1.2\arcsec -1$,$R_{e,out}/0.1\arcsec -1$].
For $0\arcsec \le R_{e,out} < 1.2\arcsec$, the size range spanned by our sample (see the text for more details), there are no significant systematics in the $R_e$ and $n$ parameters.
Using the maximum scatter for binned simulated data, we assign an error of 20\% to our measured $R_e$ and $n$.
}
	\label{fig:galfit_sim}
\end{figure*}

In Figure \ref{fig:galfit_sim}, we compare the estimated and input parameters
(magnitude: $(z_{850,out}-z_{850,in})$, effective radius: $\delta R_e = (R_{e,out}-R_{e,in})/R_{e,in}$, S\'ersic index: $\delta n = (n_{out}-n_{in})/n_{in}$) versus the input effective radius $R_{e,in}$ (left panel) and the measured effective radius $R_{e,out}$ (right panel).  In the same figure, we bin the values on the x-axis: the mean value and the scatter for each bin are shown.  For the right panel, we also show in red an histogram (arbitrary units) of the $R_e$ distribution for our real ETGs.
The green dashed lines delimit the possible $\delta R_e$ values due to the range of the simulation input values $R_{e,in}$. For example,  because of the definition of $\delta R_e$ and of $0.1\arcsec \le R_{e,in} \le 1.2\arcsec$, a galaxy with $R_{e,out}$ will necessary have a corresponding value of $\delta R_e$ within [$R_{e,out}/1.2\arcsec -1$,$R_{e,out}/0.1\arcsec -1$].

When looking at $\delta R_e$ as a function of $R_{e,in}$ (left panel), we observe that our method recovers the effective radius with no significant bias, except for large galaxies ($R_{e,in} \gtrsim 1$\arcsec), where it slightly underestimates (by $\sim 5\%$) the radius, because a significant part of the light is lost in the background noise.  When looking at $\delta R_e$ as a function of $R_{e,out}$ (left panel), we again observe no significant bias except for the galaxies with $R_{e,out} \ge 1.2$\arcsec, which have their size overestimated.  This is a direct consequence of the chosen range for $R_{e,in}$ ([0.1\arcsec,1.2\arcsec]): as the green dashed line illustrates, all our simulated galaxies with $R_{e,out} \ge 1.2$\arcsec~ can only have their size overestimated. Our real ETGs never have values of derived $R_e$ so high, as shown in the red histogram.
These correlations propagate to magnitudes and S\'ersic indexes.
	
In the range of our data ($0\arcsec \le R_{e,out} < 1.2\arcsec$), sizes and S\'ersic indexes are recovered with systematics smaller than 8\% and the magnitudes with systematics smaller than 0.08 mag, and also with a relatively small scatter.
Hence  our estimates of magnitude, $R_e$ and $n$ are well recovered, in the range covered by our observations.
Using the maximum scatter for binned simulated data, we assign an error of 20\% to our measured $R_e$ and $n$.
		
\subsection{de Vaucouleurs vs S\'ersic profile \label{sec:devauc_sersic}}

\begin{figure*}
	\begin{tabular}{cc}
	\includegraphics[width=0.5\textwidth]{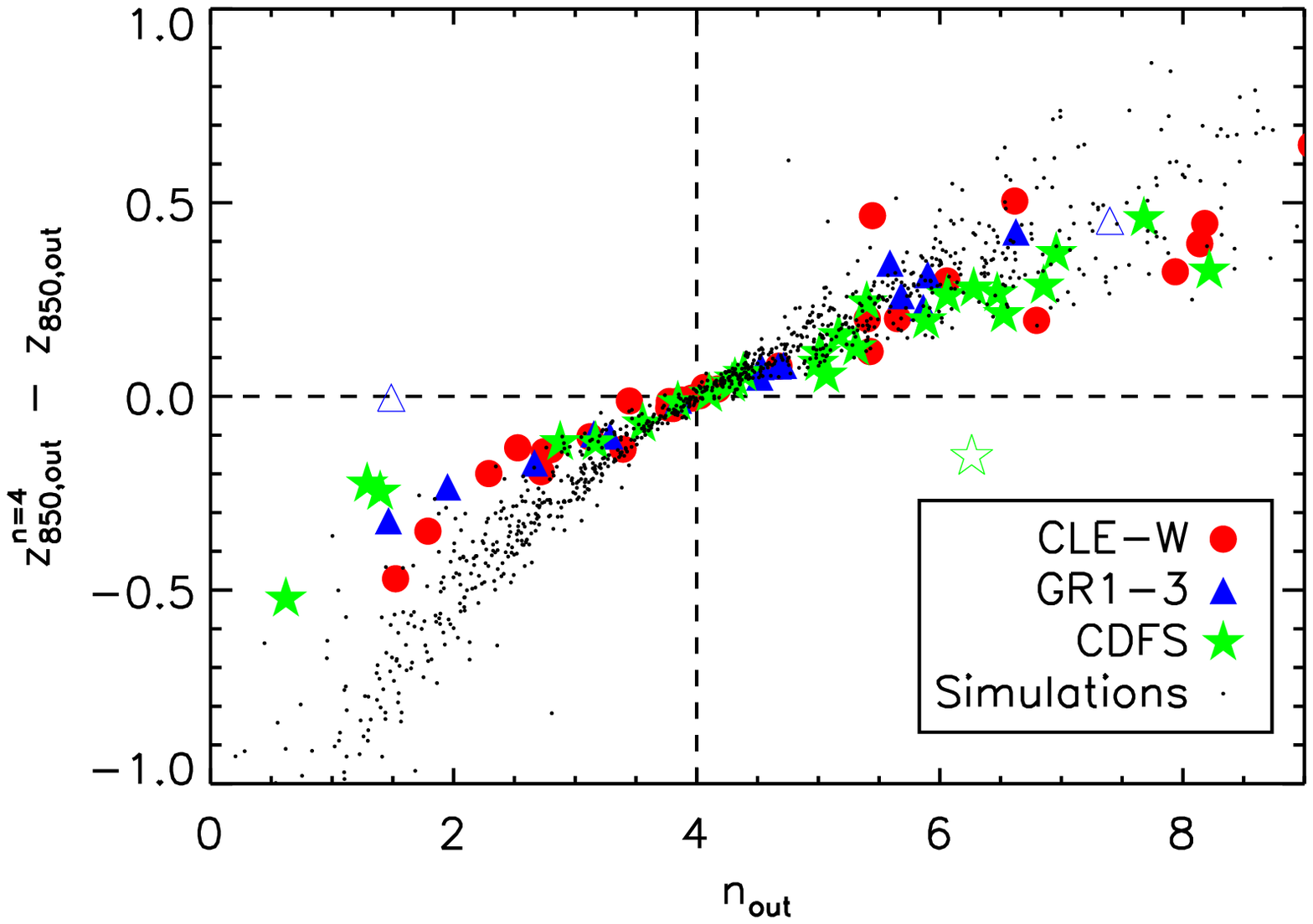}		&
	\includegraphics[width=0.5\textwidth]{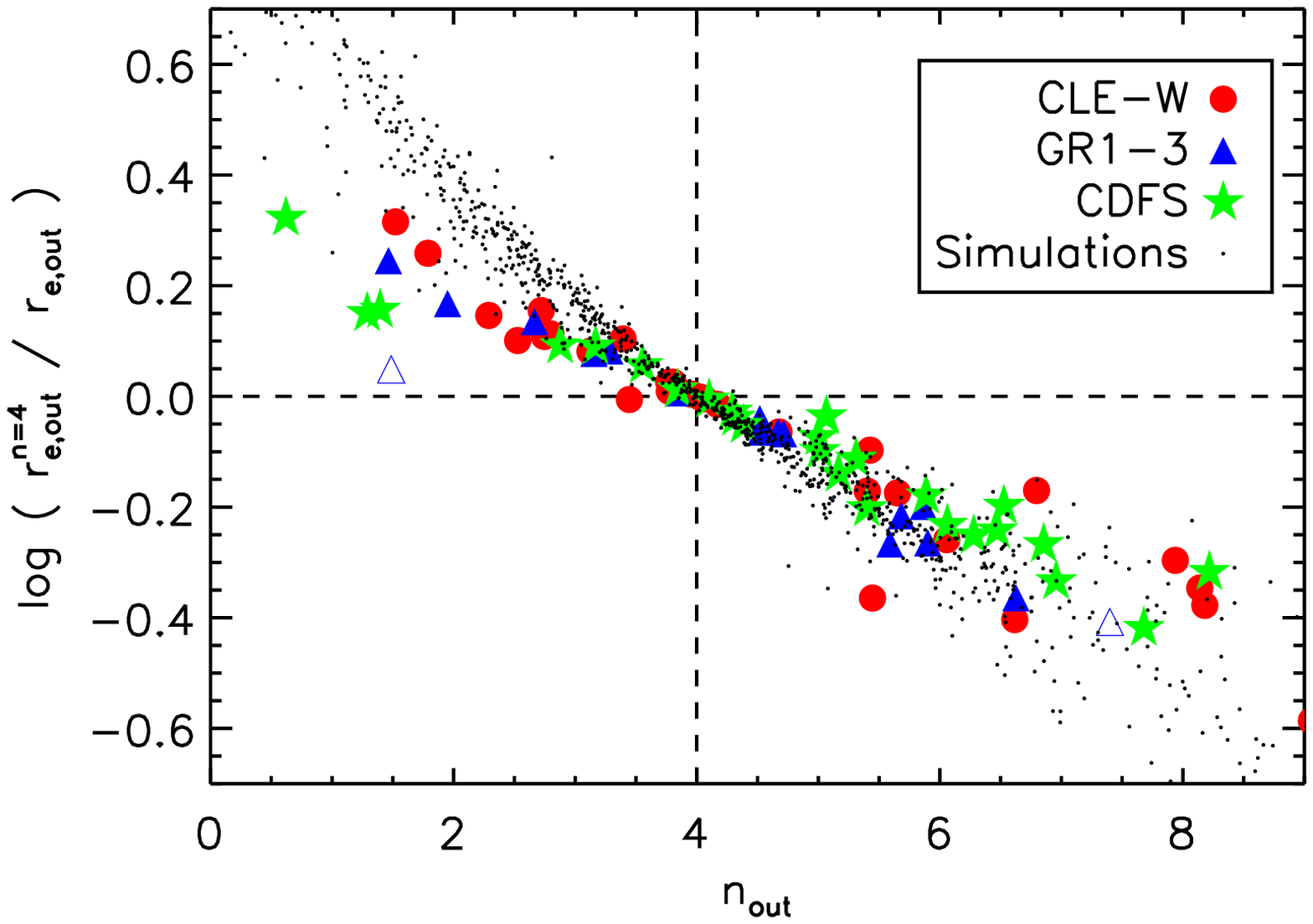}\\
	\end{tabular}
	\caption{Comparison of the estimated $z_{850}$ magnitude (\textit{left panel}) and the estimated $R_e$ (\textit{right panel}) when using a de Vaucouleurs profile (n=4) or a S\'ersic profile (n free):
real ETGs are represented with large symbols (Lynx cluster: red dots, Lynx group: blue triangles, CDF-S: green stars) and simulations with black dots.
ETGs with non robust structural parameter estimates are plotted as empty symbols.
For both magnitude and $R_e$, we observe a systematic bias depending on the S\'ersic index $n$.}
	\label{fig:galfit_vaucouleurs}
\end{figure*}

To reduce uncertainties in the fit, we tried reducing the number of free parameters by using a fixed \citet{de-vaucouleurs48} profile ($n = 4$).
We estimated how such a fit would be reliable as compared to a S\'ersic profile.  
We perform our fits again (real ETGs and simulations) with the same method, but this time with a de Vaucouleurs profile.
We then compare the results with those obtained with a S\'ersic profile ($n$ free) in Figure \ref{fig:galfit_vaucouleurs}.
The left panel represents $z_{850,out}^{n=4}-z_{850,out}$ and the right panel log($r_{e,out}^{n=4}/r_{e,out}$) as a function of the output S\'ersic index $n_{out}$.
Real ETGs are represented with large symbols (Lynx cluster: red dots, Lynx group: blue triangles, CDF-S: green stars) and simulations with black dots.
ETGs with non robust structural parameter estimates are plotted as empty symbols.
Real ETGs measurements and simulations show the same trend:
measuring the size by assuming a de Vaucouleurs profile introduces a significant bias, which depends on the S\'ersic index of the ETG.
Those results are in qualitative agreement with those of \citet{donofrio08} and \citet{taylor10}.
We will use size and $n$ estimates obtained with a S\'ersic profile hereafter.
The tables in Appendix A present sizes and surface brightnesses derived with both S\'ersic and de Vaucouleurs profiles.
We present in Figure \ref{fig:nser_hist} in Appendix B the S\'ersic index distributions for our sample.
We remark that the presence of few ETGs with small S\'ersic indexes is not unexpected, as they have been visually selected through their morphology.

\section{Kormendy relation (KR) \label{sec:kr}}

A powerful tool to investigate the ETG evolution and constrain the underlying processes is the Fundamental Plane \citep{djorgovski87,dressler87}, which is a scaling relation between the effective radius $R_e$, the mean surface brightness $\langle \mu \rangle_e$ and the central velocity dispersion $\sigma_0$.
As obtaining velocity dispersions of ETG at z $\sim$ 1 is observationally expensive, many studies have focused on the projection of the Fundamental Plane along the velocity dispersion axis, known as the Kormendy Relation (KR) \citep{kormendy77}:
\begin{equation}
\langle \mu \rangle_e = \alpha + \beta \times log \: R_e,
\end{equation}
where $R_e$ is in kpc.
The value of $\alpha$ depends on the photometric band and on the redshift.
The slope $\beta$ has been found to be constant out to z = 0.64 \citep{la-barbera03}.

We convert our $z_{850}$ magnitudes in the $B$--band rest--frame, $B_{z_0}$,  in order to derive the $B$--band rest-frame surface brightness $\langle \mu^B \rangle_e$.
We use the index $z_0$ to refer to the rest-frame and $z_{obs}$ to refer to the observed frame.
To estimate $B_{z_0}$, we use a method similar to the one used in \citet{mei09}.  We use CB07 models (choosing BC03/MA05 models changes $B_{z_0}$ by less than 0.1 mag) and consider a set of galaxies with a redshift of formation $1.8 \le z_{form} \le 7$, a solar metallicity and an exponentially declining SFH with $0.1 \le$ SFH $\tau$ (Gyr) $\le 1$.  We then linearly fit the relation between the colors ($B_{z_0} - z_{850,z_{obs}}$) and ($i_{775,z_{obs}} - z_{850,z_{obs}}$) (where $i_{775,z_{obs}}$ and $z_{850,z_{obs}}$ are the apparent magnitudes in the $i_{775}$ and $z_{850}$ bands for galaxies observed at $z=z_{obs}$).  Once this relation is established, we can estimate the total $B$--band rest-frame magnitude $B_{z_0}$ from the full measured apparent magnitude in the $i_{775}$ and $z_{850}$ bands (published in R11).  Eventually, we transform this magnitude into mean surface brightness by averaging half of the total flux on the surface within $R_e$ and correct for the cosmological dimming $(1+z_{obs})^4$:
\begin{equation}
\langle \mu^B \rangle_e = B_{z_0} +2.5 \times log(2 \pi R_e^2) - 10 \times log(1+z_{obs}).
\end{equation}
Taking into account the different steps in estimating $\langle \mu^B \rangle_e$, we assign an uncertainty of 0.4 mag for our $\langle \mu^B \rangle_e$ estimate.

\begin{figure*}
	\includegraphics[width=\linewidth]{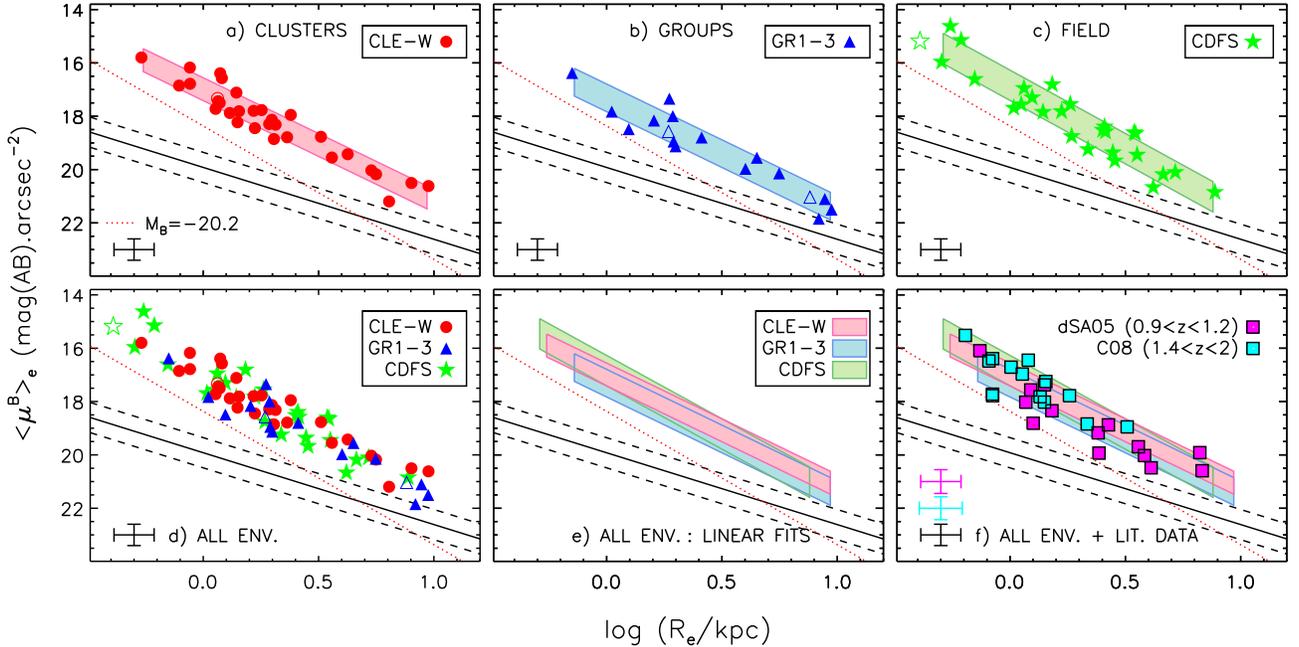}
	\caption{
\textit{Upper panels} \textit{a), b), c)}: 
Kormendy relation for our ETGs at $z \sim 1.3$ in rest-frame $B$--band.
Lynx cluster ETGs (\textit{a)}, red dots), Lynx group ETGs (\textit{b)}, blue triangles) and CDF-S ETGs (\textit{c)}, green stars).
ETGs with non robust structural parameter estimates are plotted as empty symbols.
For each environment, the colored area represent the 1$\sigma$ dispersion of the best linear fit to our data.
Typical uncertainties are represented by the cross in the lower left corner.
The black solid and dashed lines represent the local KR  \citep{jorgensen95} and the 1$\sigma$ dispersion for the Coma cluster, respectively.
The red dotted line represents a line of constant absolute magnitude $M^B_{z_0} = -20.2$ mag, corresponding to our cut in selection at $z_{850} = 24$ mag.
\textit{Lower panels} \textit{d), e), f)}: 
Kormendy relation for the three environments simultaneously.
In (\textit{d}) we show our data  and in  (\textit{e}) the areas corresponding to the 1$\sigma$ scatter of a linear fit.
In (\textit{f}): same as panel \textit{e)}, but with data from the literature; dSA05 corresponds to the sample of \citet{di-serego-alighieri05} and C08 corresponds to the sample of \citet{cimatti08} (see text).
Our Kormendy relation does not depend on the environment and is in qualitative agreement with passive evolution when comparing with $z \sim 0$ and $z \sim 1$-2 literature data.}
	\label{fig:kr}
\end{figure*}

The KR we obtain is plotted in Figure \ref{fig:kr}.  The upper panels show our KR for the three environments: Lynx cluster ETGs (left panel \textit{a)}, red dots), Lynx group ETGs (middle panel \textit{b)}, blue triangles) and CDF-S field ETGs (right panel \textit{c)}, green stars).  ETGs with non robust structural parameter estimates are plotted as empty symbols.  For each environment, the colored area represents the 1$\sigma$ dispersion of the best linear fit to our data, done through a classical chi-square error statistic minimization.
The red dotted line represents a line of constant absolute magnitude $M^B_{z_0} = -20.2$ mag, corresponding to our cut in selection at $z_{850} = 24$ mag.
The black solid and dashed lines represent the local KR: they represent the best linear fit and its 1$\sigma$ dispersion to the data measured in the $B$--band by \citet{jorgensen95} for 31 ETGs in the Coma cluster (converted to the AB magnitude system).
As \citet{jorgensen95} sizes are estimated with a de Vaucouleurs profile and we have demonstrated that this changes the size estimate (Figure \ref{fig:galfit_vaucouleurs}), we use an approach similar to \citet{la-barbera03} and exclude from this sample the three largest galaxies (log($R_e$/kpc) $\gtrsim 1$) for which the size difference between a de Vaucouleurs and a Sersic profile is likely to be significant.
In order to show that this choice of local relation does not affect our conclusions on the KR, we display in Figure \ref{fig:kr_vauc} in Appendix B a figure similar to Figure \ref{fig:kr}, but with our sizes estimated with a de Vaucouleurs profile instead of a S\'ersic profile and with including in \citet{jorgensen95} local relation the three largest galaxies.

\subsection{KR: dependence on the environment}

As shown in previous works at $z > 1$, the KR is in place at $z \sim 1.3$ in the field \citep[e.g.,][]{di-serego-alighieri05,longhetti07,cimatti08,damjanov09,saracco09} and in clusters \citep[e.g.,][]{holden05,rettura10}.
We find that, though the range in size is similar for the three environments, the distribution of cluster ETGs seems to be more concentrated towards smaller sizes.
We will come back to this point in $\S$\ref{sec:msr}. 
We then plot the data for our whole sample in the lower left panel \textit{d)} and the 1$\sigma$ dispersion around the best linear fit relations in the lower middle panel \textit{e)}.
The KRs in the three environments are in agreement, and we do not observe any dependence of the KR on the environment at $z \sim 1.3$.
\citet{rettura10} studied the KR in the field and in a cluster at $z \sim 1.2$ and found no dependence on the environment.
Our work confirms this study and extends its results to the group environment.

\subsection{KR: comparison with the local relation}

When comparing with the local KR, we observe that our relation is shifted towards brighter luminosities and the slope is steeper. This change in slope may be linked to the magnitude cut due to the depth of our $z_{850}$ image (see the line showing the depth of our ACS image in Figure \ref{fig:kr}), or be a real steepening of the KR.

\begin{figure*}
	\includegraphics[width=\linewidth]{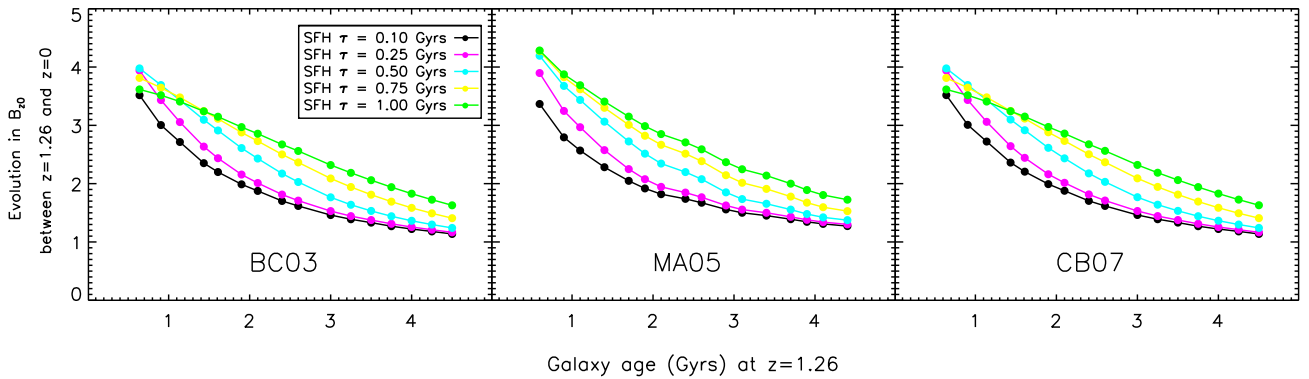}
	\caption{Theoretical luminosity evolution in rest-frame $B_{z_0}$ magnitude: 
for different SFH $\tau$, we plot the difference $B_{z_0}$($z$=1.26) - B$_{z_0}$($z$=$z_0$) predicted by the three stellar population models as a function of the age of the ETG at $z=1.26$.
Older ETGs at $z=1.26$ evolve less in $B_{z_0}$.
}
	\label{fig:lum_evol}
\end{figure*}

Stellar population models predict that the luminosity evolution depends on galaxy age and its SFH. In Figure \ref{fig:lum_evol}, we show these dependences between $z \sim 1.26$ and $z = 0$. 
We plot the luminosity evolution as a function of age at the given redshift for several exponentially declining SFHs with characteristic time SFH $\tau$ ranging from 0.1 to 1 Gyr.
The range in SFH $\tau$ encompasses the likely values for our ETGs: for all models (BC03/MA05/CB07), our estimated maximum SFH $\tau$ is below 1 Gyr for 90\% of our sample \citep[R11; see also ][]{rettura11}.
If evolving passively down to $z=0$, a 3 Gyr old ETG at $z=1.26$ will be 1.5-2.5 mag less luminous in $B_{z_0}$ whereas a 1 Gyr old ETG at $z=1.26$ will be 3-3.5 mag less luminous in $B_{z_0}$. Older ETGs evolve less in $B_{z_0}$.

\begin{figure*}
	\includegraphics[width=\linewidth]{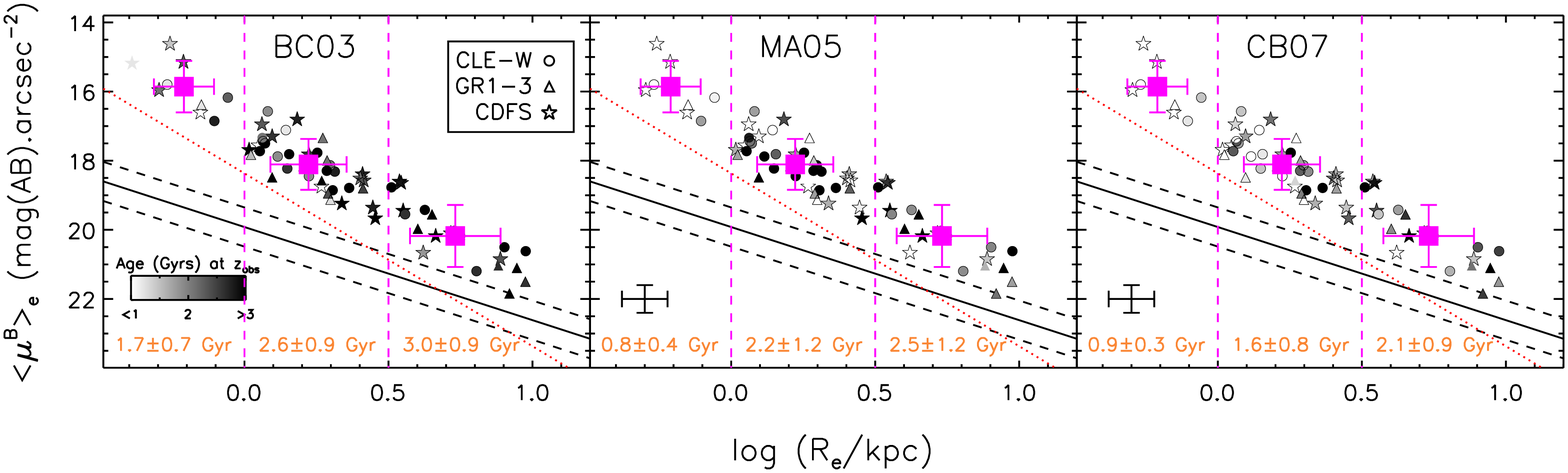}
	\caption{Kormendy relation at $z \sim 1.3$: dependence on age.
Our KR is plotted with ages estimated from three models (BC03/MA05/CB07) which are coded in gray levels.
ETGs with non robust structural parameter estimates are plotted without black outlines. 
The solid and dashed black lines and the red dotted lines are the same as in Figure \ref{fig:kr}. 
We bin our data in three size bins (vertical dashed magenta lines): for each bin, we overplot as magenta squares the mean and the 1$\sigma$ dispersion for log($R_e$/kpc) and $\langle \mu^B \rangle_e$ values.
We also report, in orange at the bottom of the figure, the mean and standard deviation of the estimated ages for each bin.
On average, large ETGs are older than smaller ones.
}
	\label{fig:kr_age}
\end{figure*}

In Figure~\ref{fig:kr_age}, we code our galaxy ages (as derived from R11, see this paper for details) in gray levels,  for the three models (BC03/MA05/CB07). Larger ETGs tend to be older.  To better visualize this trend, we bin our data in three size bins (log($R_e$/kpc) $<$ 0, 0 $\le$ log($R_e$/kpc) $<$ 0.5 and log($R_e$/kpc) $\ge 0.5$).
For each size bin, we overplot as magenta squares the mean and the 1$\sigma$ dispersion for log($R_e$/kpc) and $\langle \mu^B \rangle_e$ values.
We report, in orange at the bottom of the figure, the mean and standard deviation of the estimated ages for each bin.
Thus, we observe an age gradient in our KR, larger ETGs being on average older, because they are on average more massive \citep[see also R11; such an age gradient is observed in the local Universe, see for instance][]{shankar10}.
Under the assumption of only passive evolution in luminosity and according to stellar population models, this age gradient should lead to a steepening of the slope of the KR with increasing redshift, in qualitative agreement with what we observe.

\subsection{KR: comparison with literature data at $z \sim 1$-2}

On the panel \textit{f)} of Figure \ref{fig:kr}, we overplot as squares with black outlines the KRs published for ETGs at $z_{spec} \sim 1$-2 .
The sample from \citet{di-serego-alighieri05} (in magenta) is composed of 16 field ETGs (from the K20 survey, selected according to their spectra) at $0.9 \le z_{spec} \le 1.2$ (we removed the two ETGs at $z_{spec} \sim 0.67$).
The size is estimated by using \textsc{Gim2d} \citep{simard02} and fitting a S\'ersic profile on \textit{HST}/ACS $z_{850}$-band and \textit{VLT}/FORS-1 $z$-band images.
The sample from \citet{cimatti08} (in light blue) is composed of 13 passive galaxies (6 in the field and 7 in a cluster-like structure) selected from the GMASS project, mainly ETGs, with $1.4 \le z_{spec} \le 2$.
The size is estimated using \textsc{Galfit} by fitting S\'ersic profiles to \textit{HST}/ACS $z_{850}$-band images.
For those two samples, the radius is the circularized effective radius.  We converted the surface brightness to the AB magnitude system for the sample of  \citet{di-serego-alighieri05}.

From this comparison, we can observe two facts.  Firstly, we observe that our KR is broadly consistent with those two studies.  The KR from \citet{di-serego-alighieri05}, observed at lower redshifts, is slightly shifted towards fainter luminosities and the KR from \citet{cimatti08}, observed at higher redshifts, is lying on the higher luminosity side of our KR.  Thus, putting together those three KRs, we see a shift of the KR towards bright luminosities with increasing redshift, qualitatively consistent with passive luminosity evolution.

Secondly, looking at the range in size, we notice that the sample of  \citet{di-serego-alighieri05} lacks galaxies smaller than 1 kpc, even if comparable to our sample in a $K_s$-band limit magnitude, and is comparable to our sample for large galaxies.
The sample of \citet{cimatti08} lacks galaxies larger than $\sim$3 kpc and is comparable to our sample for small sizes.
We remark that the observed lack of small/large galaxies in those two samples is not a selection effect due to the depth of the images, which would produce a cut along a line parallel to the red dotted line (\citet{di-serego-alighieri05} limiting magnitude is about $M^B_{z_0} = -20.1$ mag and \citet{cimatti08} data are deep enough to detect a galaxy at $M^B_{z_0} = -20.2$ mag).
Our sample ranges a larger interval in size that both other samples.


\section{Mass-size relation (MSR) \label{sec:msr}}

\begin{figure*}
	\includegraphics[width=\linewidth]{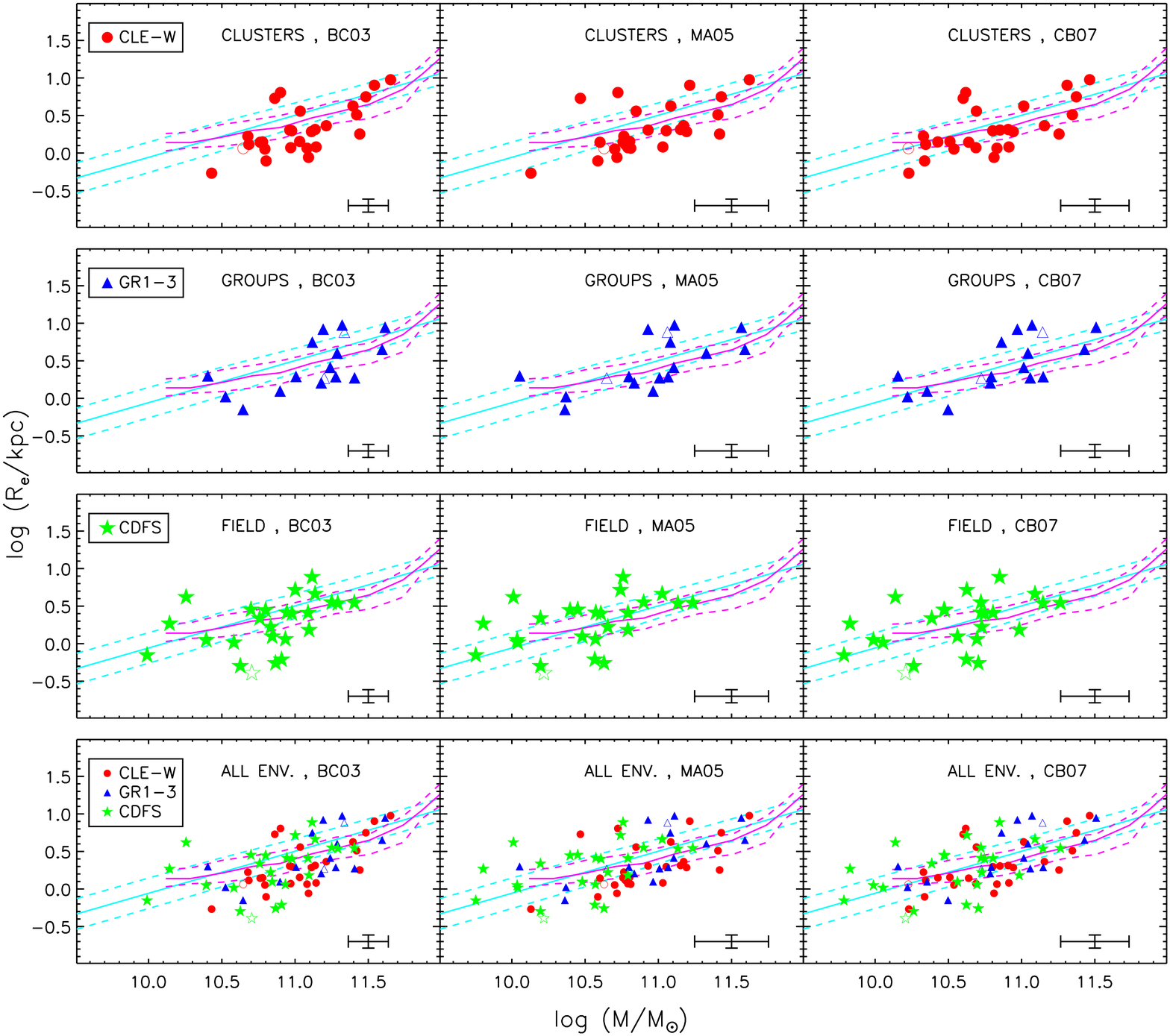}
	\caption{Mass-size relation at $z \sim 1.3$ derived with the three stellar population models (BC03/MA05/CB07), split by environments:
from upper to lower panels, Lynx cluster ETGs (red dots), Lynx group ETGs (blue triangles), CDF-S ETGs (green stars) and all environments simultaneously.
ETGs with non robust structural parameter estimates are plotted as empty symbols.
Magenta (resp. cyan) solid and dashed lines represent the local relation of \citet{valentinuzzi10} \citep[resp.][]{shen03} and its 1$\sigma$ dispersion.
Typical uncertainties are represented by the cross in the right left corner.
When quantifying size evolution, one has to be careful to the model and local MSR used (see Table \ref{tab:size_ratio}).
Most of cluster and group ETGs lie below local MSRs, whereas field ETGs are in agreement with local MSRs (except with BC03 models, for which more field ETG lie below the local MSRs, see text for discussion).
}
	\label{fig:msr}
\end{figure*}

In Figure \ref{fig:msr}, we plot our MSR, derived using the three stellar population models (BC03/MA05/CB07), and splitting our sample by environment: we display from upper to lower panels, Lynx cluster ETGs (red dots), Lynx group ETGs (blue triangles), CDF-S ETGs (green stars) and all environments simultaneously.
The solid and dashed lines represent the local MSR scaled to a Salpeter IMF, and its 1$\sigma$ relation, respectively.
The local MSR established by \citet{shen03} with sizes estimated in $z$ band for SDSS galaxies selected according to their S\'ersic index ($n \ge 2.5$) is in cyan.
The local MSR established by \citet{valentinuzzi10} with sizes estimated in $V$ band for WINGS cluster galaxies that were morphologically selected to be ETGs is in magenta.
The difference in the rest-frame used to estimate sizes would shift the local MSRs towards larger sizes \citep[around 10\%-15\% according to ][]{bernardi03a}.
We compare our sample with \citet{valentinuzzi10}, because both select ETGs from a morphologically classification.
All our results do not change when using \citet{shen03} local MSR, which is widely used in the literature.

\subsection{MSR: dependence on the environment}
	
\begin{figure*}
	\includegraphics[width=\linewidth]{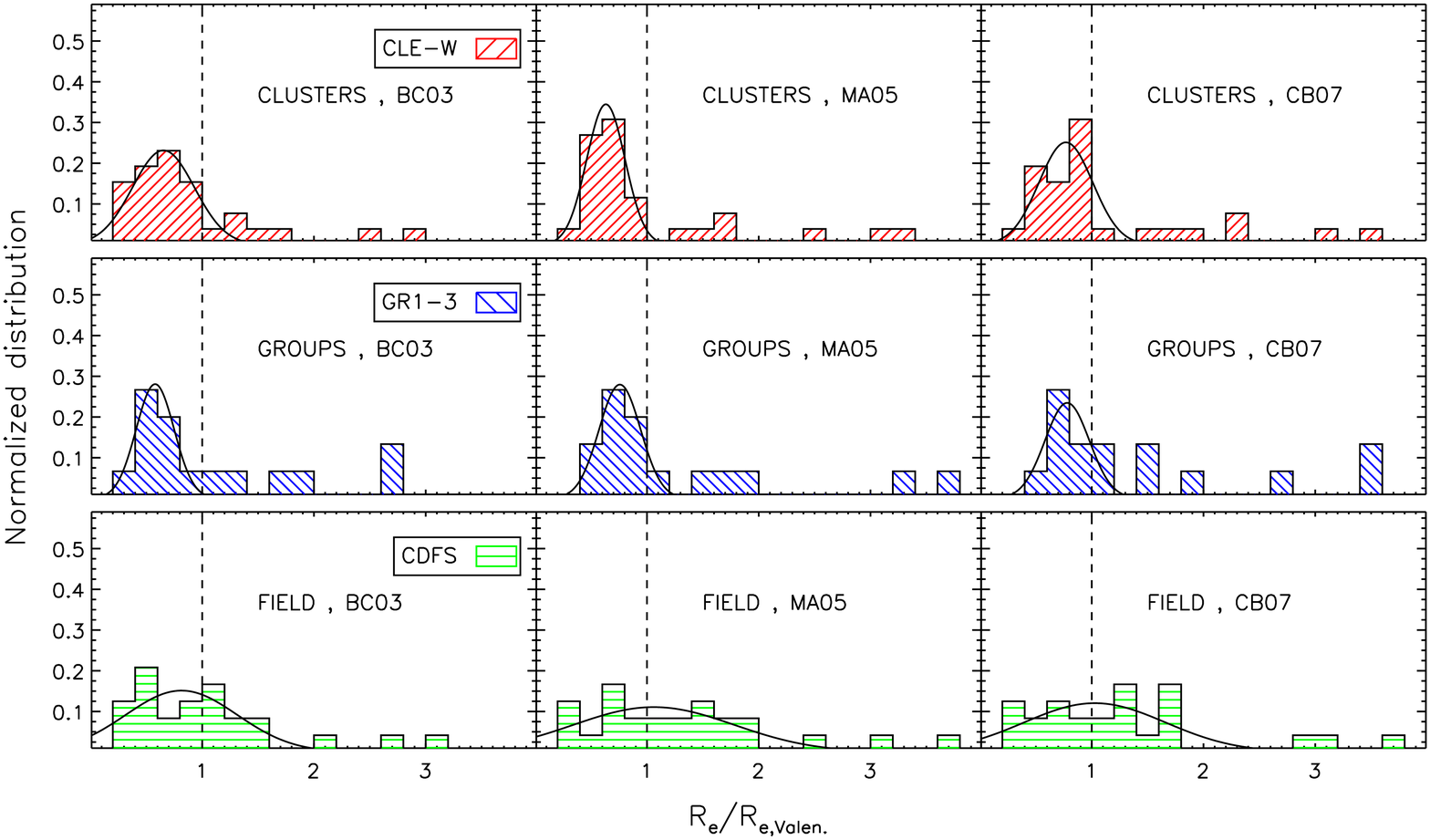}
	\caption{Size ratio $R_e/R_{e,Valen}$ normalized distributions derived with the three stellar population models (BC03/MA05/CB07), split by environments:
Lynx cluster ETGs (\textit{upper panels}, red tilted lines), Lynx group ETGs (\textit{middle panels}, blue tilted lines) and CDF-S ETGs (\textit{lower panels}, green horizontal lines).
$R_e/R_{e,Valen}$ represents the ratio between the size of our ETGs and the one predicted by the local MSR of \citet{valentinuzzi10}.
The black solid line represents the best-fit gaussian to the distributions.
The black dashed line represents the locus of $R_e/R_{e,Valen} = 1$.
For MA05/CB07 models, field ETG population is in agreement with the local MSR, whereas cluster/group ETG population lie below the local MSR.
}
	\label{fig:msr_hist}
\end{figure*}

	We plot in Figure \ref{fig:msr_hist}, for the three environments and the three models, the normalized distributions of the size ratio $R_e/R_{e,Valen.}$, which represents the ratio between the size of our ETGs and the one predicted by the local MSR of \citet{valentinuzzi10} at similar masses.
For each histogram, we overplot with a black solid line the best-fit gaussian to the distribution, obtained through a non-linear least-squares fit.
As found in previous studies at $z \sim 1$-2, our sample presents a significant number of ETGs having small radii compared to the local ones of similar mass.
However, the precise number of such ETGs and the value of size ratios depend on the model (and the local MSR used as a reference).
We display in Table \ref{tab:size_ratio} the mean and standard deviation corresponding to the gaussian fit.
For qualitative comparison, we also display in this Table the corresponding values when comparing our sizes to \citet{shen03} local MSR.

\begin{deluxetable*}{l c c c c c c}
	\tablewidth{\linewidth}
	\tablecaption{Size Ratios as a Function of Stellar Population Model, Local MSR, and Environment \label{tab:size_ratio}}
	\tablehead{ Environment
				&	\multicolumn{3}{c}{$R_e/R_{e,Valen}$}							&	\multicolumn{3}{c}{$R_e/R_{e,Shen.}$}	\\
				&	\multicolumn{3}{c}{}											&	\multicolumn{3}{c}{}	\\
				&	BC03		&	MA05		&	CB07					&	BC03		&	MA05		&	CB07}
	\startdata
				&				&				&							&				&				&				\\
CLE-W			&	0.7 $\pm$ 0.3	&	0.6 $\pm$ 0.2	&	0.8 $\pm$ 0.2				&	0.5 $\pm$ 0.2	&	0.5 $\pm$ 0.1	&	0.7 $\pm$ 0.2	\\
GR1-3			&	0.6 $\pm$ 0.2	&	0.8 $\pm$ 0.2	&	0.8 $\pm$ 0.2				&	0.6 $\pm$ 0.2	&	0.7 $\pm$ 0.2	&	0.8 $\pm$ 0.2	\\
CDF-S			&	0.8 $\pm$ 0.5	&	1.1 $\pm$ 0.7	&	1.0 $\pm$ 0.6				&	0.7 $\pm$ 0.3	&	1.1 $\pm$ 0.6	&	1.0 $\pm$ 0.5	\\
				&				&				&							&				&				&				\\
\hline
				&				&				&							&				&				&				\\
All				&	0.7 $\pm$ 0.4	&	0.7 $\pm$ 0.2	&	0.8 $\pm$ 0.4				&	0.6 $\pm$ 0.3	&	0.6 $\pm$ 0.2	&	0.8 $\pm$ 0.3	\\
	\enddata
\end{deluxetable*}

Despite the dependence on the model (and on the local MSR), there is a general trend : most of the cluster and group ETGs lie below local MSRs.  We observe in Figure \ref{fig:msr_hist} that cluster and group ETG size ratios are mostly below 1 with a narrow distribution that peaks around 0.6-0.8, whereas field ETG size ratios have a more widespread distribution, peaking around 0.7-1.1.  The values in Table \ref{tab:size_ratio} confirm this point, i.e. that {\it at a given mass, ETGs in denser environments tend to have smaller sizes at $z\sim1.3$ than in the local Universe}.
From a Kolmogorov-Smirnov (Kuiper) statistical test, the cluster and field samples for ETGs with masses $M < 10^{11} M_{\sun}$ do not have the same size ratio distributions, at 85\% (90\%) and 90\% (95\%) using MA05 and CB07 models, respectively. Using BC03 stellar population models, on the other hand, the null hypothesis (the cluster and field samples are taken from the same statistical distribution)  cannot be rejected (rejected at only 40\% and 60\% confidence for a Kolmogorov-Smirnov and Kuiper test, respectively), in this mass range, and it is rejected at 86\% and 90\%, respectively, on the entire mass range. We underline that the Kuiper test is more sensitive to the shape of the distribution than the Kolmogorov-Smirnov test.

We remark that our size ratios in clusters are in agreement with previous estimates in high redshift clusters \citep{rettura10,strazzullo10}.
We know that ETGs with emission lines have higher size ratios \citep[e.g.][]{toft07,zirm07,williams10}.
Even when we remove them  \citep[][Holden, Nakata, private communication]{salimbeni09} our distributions remain similar as in Figure \ref{fig:msr_hist}, as it can be seen in Figure \ref{fig:msr_hist_noem} in Appendix B.

	The field ETG population is approximately equally divided with ETGs above and below the local MSRs, except when using the BC03 models, in which case a majority lie below the local MSRs.  This can be explained by the underestimate of the TP-AGB phase by BC03 models.  As explained in previous works \citep[e.g.][R11]{maraston06,conroy10}, this underestimate leads to an artificial increase of the estimated age and mass of galaxies with ages $\sim$1-2 Gyr (see Figure 5 \& 6 of R11).  Such an ETG will be estimated with an older age and a greater mass: it will be shifted towards the more massive area in Figure \ref{fig:msr} and thus will more likely lie below the local MSRs.  This effect of BC03 models is more obvious for our CDF-S sample, because this sample contains more ETGs with ages $\sim$1-2 Gyr (see R11). 

	In R11, we underlined that our CDF-S sample might be biased against low-mass/passive ETGs, because their low luminosity and the lack of emission line prevent to derive reliable spectroscopic redshift. We have tested that our conclusions for the field sample do not depend on this potential bias. We have selected a new sample of GOODS/CDF-S ETGs, using photometric redshifts from \citet{santini09} and we obtain a MSR that again shows a distribution similar to the local. The selection criteria are described in Appendix B and the results are in Figure \ref{fig:msr_cdfs_test}.

	If on the other hand, we are missing massive galaxies, this would not change the overall mass-size distribution, which clearly shows to be similar to the local one at all masses and will not be changed significantly by rare massive galaxies.

\subsection{Size ratio versus redshift of formation/stellar mass}

\begin{figure*}
	\includegraphics[width=\linewidth]{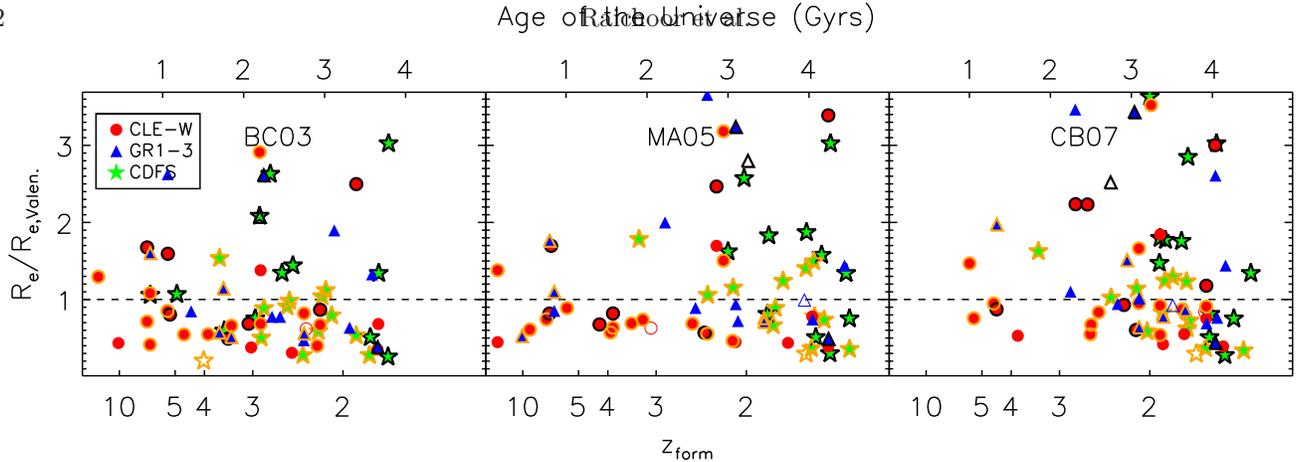} \\
	\caption{Size ratio versus redshift of formation, derived with the three stellar population models (BC03/MA05/CB07):
Lynx cluster ETGs are red disks, Lynx group ETGs are blue triangles and CDF-S ETGs are green stars.
ETGs with non robust structural parameter estimates are plotted as empty symbols.
We also mark with thick black outline the ETGs known to have emission lines and with thick orange outline the ETGs known to be passive.
The black dashed line represents the locus of $R_e/R_{e,Valen} = 1$.
For MA05/CB07 models and cluster/group environments, ETGs with small size ratios have equally distributed $z_{form}$ and ETGs with high size ratios are younger ETGs ($z_{form} \lesssim 3$).
For the field, our sample do not show old galaxies (see also R11) probably because it does not cover large areas. However, again, field galaxies cover a larger distribution in sizes.
}
	\label{fig:reff_ratio}
\end{figure*}

We now check the dependence of the size ratio on the redshift of formation.
We plot in Figure \ref{fig:reff_ratio} the size ratio $R_e/R_{e,Valen.}$ as a function of the redshift of formation $z_{form}$. 
For indication, we also mark with thick black outline the ETGs known to have emission lines and with a thick orange outline the ETGs known to be passive \citep[][Holden, Nakata, private communication]{salimbeni09}.
As already observed in the literature \citep[e.g.][]{valentinuzzi10,williams10}, when looking at our results with MA05/CB07 models, we observe that passive/quiescent ETGs tend to have small size ratios, whereas line-emitting/star-forming ETGs tend to have larger size ratios.

	If we consider the plots with MA05/CB07 models (for which the redshift of formation is more robust because of the better modeling of the TP-AGB stellar phase), most of cluster and group ETGS  lie below the local MSRs, as already shown in Section~5.1. We also observe that galaxies with small size ratios do not have a preferred redshift of formation and, on the other hand, there is a deficit of ETGs with $z_{form} \gtrsim 3$ and a high size ratio ($R_e/R_{e,Valen.} \gtrsim 2$).

\begin{figure*}
	\includegraphics[width=\linewidth]{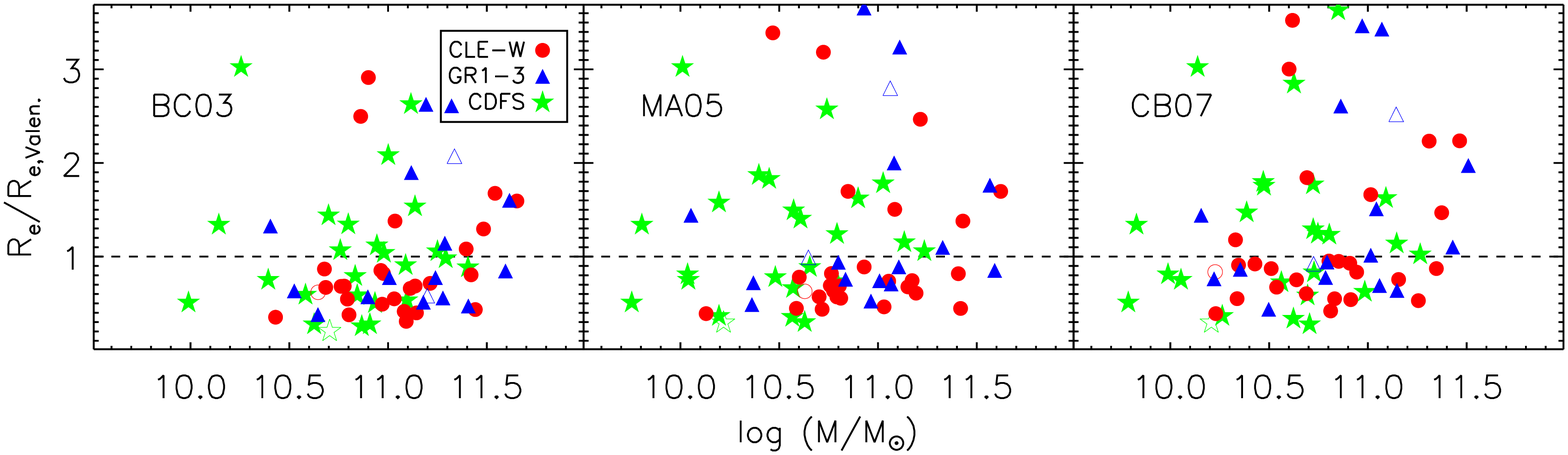}
	\caption{Size ratio versus stellar mass, derived with the three stellar population models (BC03/MA05/CB07):
symbols are as in Figure \ref{fig:reff_ratio}.
We do not observe any clear dependence of the size ration on the stellar mass.
}
	\label{fig:reff_ratio_mass}
\end{figure*}

	Figure \ref{fig:reff_ratio_mass} shows the dependence of the size ratio on the stellar mass.
We do not observe any clear dependence of the size ratio on the stellar mass.
In particular, for MA05/CB07 models and all environments, ETGs with small size ratio span the whole range in mass of our sample.
Moreover, the high-mass end of our sample ($M \gtrsim 2 \times 10^{11} M_{\sun}$, see also Figure \ref{fig:msr}) is in agreement with the local MSR for all environments.

\subsection{MSR: comparison with literature data at $z \sim 1$-2}

Our results are consistent with published works of the MSR at the same redshift and with comparable mass range, i.e. $10^{10} \lesssim M/M_{\sun} \lesssim 3 \times 10^{11}$.

\citet{valentinuzzi10a}, whose sample covers masses slightly higher to those of this work up to a redshift $z \sim 0.7$, have found that the median MSR in galaxy clusters have only mildly evolved between $z \sim 0.7$ to the present (size ratio $\sim 0.5$-0.8 when comparing ETG populations).
Working with a \citet{kroupa01} IMF and on a mass-selected sample ($M_{Kroupa} \ge 4 \times 10^{10} M_{\sun}$), and defining a superdense galaxy as a galaxy with $\Sigma_{50} = \frac{0.5 \times M_{Kroupa}}{\pi R_e^2} \ge 3 \times 10^{9} M_{\sun}$ kpc$^{-2}$, these authors find that 41\% of their $z \sim 0.7$ cluster sample is made of superdense galaxies, against 17\% for $z \sim 0$ cluster sample \citep{valentinuzzi10}.
Using the very same criteria, we find that $\sim 45\%$ (BC03) or $\sim 35\%$ (MA05/CB07) of our cluster sample is made of superdense galaxies. The mass cut reduces our cluster sample to $\sim$ 15 galaxies, lowering the statistics: assuming Poissonian errors on our number of galaxies leads to an uncertainty of $\sim 20\%$ on our estimated percentages. A correct estimate should also take into account the different galaxy selection and size estimate methods.

\citet{newman10} study a sample of field spheroidals with masses $6 \times 10^{10} < M/M_{\sun} < 3 \times 10^{11}$ in the redshift range $1.1 < z_{spec} < 1.6$.
They find that galaxies less massive than $10^{11} M_{\sun}$ lie on the local MSR, and those more massive than  $10^{11} M_{\sun}$ have to grow of around twice in size.
Those results are consistent with our founding in the field.

\citet{saracco11} study field and cluster (among which Lynx members) ETGs lying at $0.9 < z_{spec} <2$.
They find that compact ETGs formed over a wide range of redshift ($2 < z_{form} < 10$) and that normal ETGs have formed at $z \lesssim 3$.
Our results are in agreement, since we find that our cluster ETGs (more compact) formed over a wide range of redshifts and field ETGs (that have younger ages) have a larger dispersion in sizes.
In our sample these different distributions seems to be linked to the environment.
They also find compact ETGs throughout all their probed mass range ($5 \times 10^{10} < M / M_{\sun} <  5 \times 10^{11}$).

\citet{van-der-wel08} studied a sample of morphologically selected ETGs in field and cluster environments at $0.8 < z_{spec} < 1.2$ and, using dynamical masses, found that ETGs have on average increase their size by a factor of 2 between $z \sim 0$ and $z \sim 1$.
This result is at apparent discrepancy with ours for field ETGs. However, these two works probe different ranges in galaxy mass: our sample spans masses of $10^{10} < M/M_{\sun} < 3 \times 10^{11}$, whereas \citet{van-der-wel08} sample includes more massive, $8 \times 10^{10} < M/M_{\sun} < 10^{12}$ (when dynamical masses are converted to Salpeter stellar masses).
In addition, \citet{van-der-wel08} sizes are estimated using de Vaucouleurs profile, which complicates a possible comparison for the few ETGs in common with similar mass (see \S.\ref{sec:devauc_sersic} and Figure \ref{fig:galfit_vaucouleurs}). The fact that \citet{van-der-wel08} do not find an environmental dependence on the evolution of the MSR and we do, probably reflects the different range in mass probed by the two works. This points to a very interesting situation since it suggests that the evolution of the MSR is mass dependent.

Recent works have found very compact galaxies
in the field at $z_{spec} \gtrsim 1.5 $  \citep[e.g.][]{daddi05,van-dokkum08,damjanov09,van-dokkum10}.
The mass range covered in these works ($M \gtrsim 2 \times 10^{11} M_{\sun}$, when converted to Salpeter stellar masses) hardly overlaps our mass range, and we thus cannot compare conclusions directly because we are sampling different ranges in mass and in redshift.

Our results are driven by galaxies with masses $M \lesssim 2 \times 10^{11} M_{\sun}$.
Our galaxies with masses $M \sim 10^{11} M_{\sun}$ follow the same trends that the entire sample: field galaxies lie on the local MSR relation, cluster galaxies show an average MSR shifted to sizes 30-50\%  smaller.
Our galaxies with masses $M \gtrsim 2 \times 10^{11} M_{\sun}$ are a few, and show a large dispersion in size: they lie on the local MSR independently of the environment (see Figure \ref{fig:msr}), but their small number does not permit us to draw conclusions on their behaviour.


\section{Conclusions \label{sec:conclusion}}

	In this work, we have studied a sample of 76  ETGs spanning a wide range of environments (cluster, group, and field) at $z \sim 1.3$, combining multi--wavelength observations of the Lynx supercluster, with data on the GOODS/CDF-S field.
We estimated the size of our ETGs by fitting a S\'ersic profile to the \textit{HST}/ACS $z_{850}$ images, which probe the rest-frame $B$-band.
Combining those sizes with stellar masses and stellar population ages derived in R11, we are able to study two crucial structural relations, the Kormendy relation and the mass-size relation, in three different environments at $z \sim 1.3$.

We obtain the following results:
\begin{enumerate}
\item The Kormendy relation, in place at $z \sim 1.3$, does not depend on the environment.
We thus confirm the result of \citet{rettura10} and extend it to the group environment.
Our results are in agreement with results in the cluster and field samples of \citet{di-serego-alighieri05} and \citet{cimatti08}.
\item Concerning the mass-size relation, for all stellar population models (BC03/MA05/CB07) and local relations \citep{shen03,valentinuzzi10}, ETGs are on average more compact in denser environments.
When comparing the MSR at high redshift with the one in the local universe, the uncertainty on the mass coming from the model used to estimate it and the choice of the local MSR can significantly influence the conclusion on the importance of the size evolution.
When using MA05/CB07 models, we find that the majority of cluster and group ETGs are below the local relations, whereas field ETGs follow a MSR similar to the local one.
From a Kolmogorov-Smirnov (Kuiper) statistical test, the cluster and field samples for galaxies with masses $M < 10^{11} M_{\sun}$ do not follow the same size ratio distribution, at 85\% (90\%) and 90\% (95\%) using MA05 and CB07 models, respectively. When using BC03 models, the two distributions do not differ.
\item When using MA05/CB07 models, we find that compact ETGs do not have a preferred redshift of formation.
Those results are in close agreement with those of \citet{saracco11}, who studied a sample of 62 ETGs with $0.9 < z_{spec} < 2$.
As concluded by these authors, the lack of dependence of the compactness on the redshift of formation is not consistent with models that predict compact galaxies to have formed at earlier times, when the Universe was more dense.
\end{enumerate}

	When we compare the MSR of cluster and group ETGs vs field ETGs, we find that, at similar masses, cluster and group ETGs are more compact than field ETGs.
On average this does not depend on cluster galaxy age.
This result is in contrast with what has been found so far for field galaxies at $z \sim 1$ at higher masses \citep[e.g.][]{van-der-wel08}, and it might be due to the different range in masses that we are probing. If this was confirmed by larger samples, it would mean that environmental effects are visible in the evolution of the MSR for ETGs with $M \lesssim 2 \times 10^{11} M_{\sun}$.

	Our results are mainly driven by galaxies with masses  $M  \lesssim 2 \times 10^{11} M_{\sun}$.
Our galaxies with masses $M \sim 10^{11} M_{\sun}$ follow the same trends that the entire sample.
Our galaxies with masses $M \gtrsim 2 \times 10^{11} M_{\sun}$ are a few, but they lie on the local MSR independently of the environment (see Figure \ref{fig:msr}); however their small number does not permit us to draw conclusions on their behaviour.
As concluded by other authors \citep{newman10,cassata11,saracco11}, the very compact galaxies at $z \sim 2$ should have gone a dramatic evolution in size to reproduce our results at $z \sim 1.3$.
This growth between $z \sim 2$ and $z \sim 1$ seems to be somehow different in cluster and field galaxies.
\citet{cassata11} have shown that, in the field, ETGs enlarge their size and increase their stellar mass by a factor of 5 between $z=2$ and $z \sim 1$.
At $z \sim1.3$, field galaxies are already on the local MSR, while cluster galaxies still have compact sizes on average \citep[see also][for a similar result at $z = 1.4$]{strazzullo10}, indicating that their size distribution still needs to be enlarged \citep[see also][]{valentinuzzi10a}.

Since in the local Universe, the ETG MSR does not depend on environment \citep{maltby10}, our results imply that an evolution in the MSR of cluster and group ETG size is required to explain current observations, while field ETGs show a MSR that is compatible with the local one.
The evolution of the MSR in dense environments might reflect either an evolution in size of the pristine population or the transformation of ETG progenitors that are not classified as ETG at $z \sim 1.3$ or the accretion of a new population of larger ETGs.

In the first case, minor dry merger events could have enlarged the size of the ETG population. 
In the second case, compact ETGs might have not had much evolution, but a new population of larger ETGs could have been formed by not-ETG progenitors or accreted in dense environments at $z < 1$ \citep{valentinuzzi10a}.
This new population might not have been observed at $z \sim 1.3$ because its progenitors are not ETGs at that time.
These galaxies might be disk galaxies that have evolved from a large bulge spiral population or galaxy mergers \citep[e.g.,][]{postman05,mei06a,poggianti06,valentinuzzi10a,mei11}.
For instance, according to semi-analytic models, the ETG population at $z \sim 1$ in dense environments contain less than $\sim 70\%$ of the stellar mass which ends up in ETGs at $z \sim 0$ \citep{kaviraj09}.

On the other hand, these results pose some challenges to current state-of-the-art galaxy evolution models that predict a nearly mass-independent size for ETGs \citep[e.g.][]{shankar11} and, we have checked, nearly independent of environment.
More detailed theoretical work is required to fully understand all the processes at work that can affect galaxy sizes.
This is clearly beyond the scope of the present work and will be the subject
of future efforts.

\acknowledgments
ACS was developed under NASA contract NAS 5-32865.
This research has been supported by the NASA HST grant GO-10574.01-A, and Spitzer grant for program 20694.
The Space Telescope Science Institute is operated by AURA Inc., under NASA contract NAS5-26555. Some of the data presented herein were obtained at the W. M. Keck Observatory, which is operated as a scientific partnership among the California Institute of Technology, the University of California and the National Aeronautics and Space Administration.
The Observatory was made possible by the generous financial support of the W. M. Keck Foundation.
The authors recognize and acknowledge the very significant cultural role and reverence that the summit of Mauna Kea has always had within the indigenous Hawaiian community.
We are most fortunate to have the opportunity to conduct observations from this mountain.
Some data were based on observations obtained at the Gemini Observatory, which is operated by the Association of Universities for Research in Astronomy, lnc., under a cooperative agreement with the NSF on behalf of the Gemini partnership: the National Science Foundation (United States), the Science and Technology Facilities Council (United Kingdom), the National Research Council (Canada), CONlCYT (Chile), the Australian Research Council (Australia), Ministrio da Cincia e Tecnologia (Brazil) and Ministerio de Ciencia, Tecnologa e lnnovacin Productiva (Argentina),Gemini Science Program ID: GN-2006A-Q-78.
We thank the anonymous referee for a careful reading of the manuscript. A.R. thanks A. Graham for useful comments.

{\it Facilities:} \facility{HST (ACS)}, \facility{Spitzer (IRAC)}, \facility{KPNO:2.1m (FLAMINGOS)}, \facility{Hale (COSMIC)}, \facility{Keck:I (LRIS)}, \facility{Gemini:Gillett (GMOS)}


\appendix

	
\section{Appendix A: Lynx and CDF-S ETG structural parameters \label{sec:tables}}

Lynx (resp. CDF-S) ETG structural parameters are presented in Tables \ref{tab:cat_lynx_cl} and  \ref{tab:cat_lynx_gr} (resp. Table \ref{tab:cat_cdfs})

\begin{deluxetable}{l c c c c c c}
	\setlength{\tabcolsep}{0.15in}
	\tablecolumns{7}
	\tablecaption{Lynx cluster ETG structural parameters and surface brightness. \label{tab:cat_lynx_cl}}
	\tablehead{
		\colhead{ID} & 
		\colhead{R.A.} &
		\colhead{DEC.} &
		\colhead{$n$} &
		\colhead{$R_e$}	& 
		\colhead{$R_e$}	& 
		\colhead{$\langle \mu^B \rangle_e$} \\
		\colhead{} & 
		\colhead{(J2000)} & 
		\colhead{(J2000)} & 
		\colhead{} & 
		\colhead{(arcsec)} & 
		\colhead{(kpc)} & 
		\colhead{(AB mag.arcsec$^{-2}$)}\\
	}
	\startdata
\cutinhead{Lynx Cluster E ($\langle z \rangle = 1.261$)}
4945 & 08 48 49.99 & +44 52 01.78 & 6.1 & 0.67 & 5.61 & 20.2 \\
 &  &  & 4 & 0.37 & 3.12 & 18.9 \\
6229 & 08 48 55.90 & +44 51 54.99 & 9.1 & 0.77 & 6.40 & 21.2 \\
 &  &  & 4 & 0.20 & 1.66 & 18.3 \\
6090 & 08 48 56.64 & +44 51 55.76 & 2.3 & 0.14 & 1.16 & 17.4 \\
 &  &  & 4 & 0.19 & 1.63 & 18.2 \\
5355 & 08 48 57.66 & +44 53 48.69 & 5.4 & 0.16 & 1.30 & 17.9 \\
 &  &  & 4 & 0.13 & 1.04 & 17.4 \\
8713 & 08 48 57.85 & +44 50 55.32 & ... & ... & ... & ... \\
5817 & 08 48 57.91 & +44 51 52.25 & 7.9 & 0.23 & 1.93 & 18.3 \\
 &  &  & 4 & 0.12 & 0.98 & 16.8 \\
5634 & 08 48 58.53 & +44 51 33.25 & 3.8 & 0.51 & 4.23 & 19.4 \\
 &  &  & 4 & 0.54 & 4.49 & 19.6 \\
5693 & 08 48 58.60 & +44 51 57.21 & 1.5 & 0.14 & 1.19 & 16.4 \\
 &  &  & 4 & 0.28 & 2.36 & 17.9 \\
5680 & 08 48 58.63 & +44 51 59.46 & 1.8 & 0.20 & 1.65 & 17.8 \\
 &  &  & 4 & 0.36 & 2.97 & 19.1 \\
5794 & 08 48 58.67 & +44 51 56.97 & 8.2 & 0.29 & 2.40 & 18.0 \\
 &  &  & 4 & 0.12 & 0.96 & 16.0 \\
8495 & 08 48 58.93 & +44 50 33.77 & 3.4 & 0.10 & 0.88 & 16.8 \\
 &  &  & 4 & 0.10 & 0.85 & 16.7 \\
5748 & 08 48 58.95 & +44 52 10.90 & 4.1 & 0.21 & 1.79 & 17.8 \\
 &  &  & 4 & 0.21 & 1.74 & 17.7 \\
5689 & 08 48 59.10 & +44 52 04.64 & 4.2 & 0.20 & 1.67 & 18.4 \\
 &  &  & 4 & 0.19 & 1.62 & 18.4 \\
5876 & 08 48 59.72 & +44 52 51.28 & 5.7 & 0.39 & 3.24 & 18.8 \\
 &  &  & 4 & 0.26 & 2.18 & 17.9 \\
5602 & 08 49 00.32 & +44 52 14.39 & 4.7 & 0.24 & 1.98 & 18.1 \\
 &  &  & 4 & 0.21 & 1.71 & 17.8 \\
8662 & 08 49 01.07 & +44 52 09.65 & 3.8 & 0.24 & 2.03 & 18.9 \\
 &  &  & 4 & 0.26 & 2.14 & 19.0 \\
8041 & 08 49 01.52 & +44 50 49.73 & 2.7 & 0.17 & 1.39 & 17.1 \\
 &  &  & 4 & 0.21 & 1.78 & 17.7 \\
8625 & 08 49 03.31 & +44 53 04.12 & 3.8 & 0.06 & 0.54 & 15.8 \\
 &  &  & 4 & 0.07 & 0.55 & 15.8 \\
7653 & 08 49 04.52 & +44 50 16.42 & 3.9 & 0.09 & 0.79 & 16.9 \\
 &  &  & 4 & 0.10 & 0.80 & 16.9 \\
8047 & 08 49 05.34 & +44 52 03.79 & 5.4 & 0.64 & 5.36 & 20.0 \\
 &  &  & 4 & 0.43 & 3.62 & 19.2 \\
7475 & 08 49 05.96 & +44 50 37.00 & 4.0 & 0.10 & 0.88 & 16.2 \\
 &  &  & 4 & 0.10 & 0.87 & 16.2 \\
\cutinhead{Lynx Cluster W ($\langle z \rangle = 1.273$)}
1745 & 08 48 29.71 & +44 52 49.68 & 2.8 & 0.17 & 1.41 & 18.2 \\
 &  &  & 4 & 0.21 & 1.79 & 18.8 \\
1486 & 08 48 31.72 & +44 54 42.95 & 6.8 & 0.14 & 1.13 & 17.7 \\
 &  &  & 4 & 0.09 & 0.78 & 16.9 \\
1794 & 08 48 32.78 & +44 54 07.22 & 3.1 & 0.17 & 1.43 & 17.8 \\
 &  &  & 4 & 0.21 & 1.72 & 18.2 \\
1922 & 08 48 32.99 & +44 53 46.69 & 2.5 & 0.14 & 1.20 & 16.6 \\
 &  &  & 4 & 0.18 & 1.49 & 17.0 \\
1525 & 08 48 33.01 & +44 55 11.92 & 8.1 & 0.43 & 3.61 & 19.5 \\
 &  &  & 4 & 0.20 & 1.65 & 17.8 \\
1962 & 08 48 33.04 & +44 53 39.75 & 5.4 & 0.14 & 1.18 & 17.5 \\
 &  &  & 4 & 0.05 & 0.44 & 15.4 \\
2094 & 08 48 34.08 & +44 53 32.32 & 2.7 & 0.28 & 2.31 & 18.8 \\
 &  &  & 4 & 0.39 & 3.26 & 19.5 \\
2343 & 08 48 35.98 & +44 53 36.12 & 3.4 & 1.13 & 9.47 & 20.6 \\
 &  &  & 4 & 1.44 & 12.00 & 21.1 \\
2195 & 08 48 36.17 & +44 54 17.30 & 6.6 & 0.96 & 7.99 & 20.5 \\
 &  &  & 4 & 0.38 & 3.19 & 18.5 \\
2571 & 08 48 37.08 & +44 53 34.05 & 4.0 & 0.25 & 2.06 & 18.3 \\
 &  &  & 4 & 0.25 & 2.08 & 18.3 \\
\enddata
\tablecomments{$R_e$ denotes the circularized effective radius. Uncertainties on $R_e$ are of 20\%, uncertainties on $n$ are of 20\% and uncertainties on $\langle \mu^B \rangle_e$ are of 0.4 magnitudes. We did not report parameters considered as non robust.}
\end{deluxetable}

\begin{deluxetable}{l c c c c c c}
	\setlength{\tabcolsep}{0.15in}
	\tablecolumns{7}
	\tablecaption{Lynx group ETG structural parameters and surface brightness. \label{tab:cat_lynx_gr}}
	\tablehead{
		\colhead{ID} & 
		\colhead{R.A.} &
		\colhead{DEC.} &
		\colhead{$n$} &
		\colhead{$R_e$}	& 
		\colhead{$R_e$}	& 
		\colhead{$\langle \mu^B \rangle_e$} \\
		\colhead{} & 
		\colhead{(J2000)} & 
		\colhead{(J2000)} & 
		\colhead{} & 
		\colhead{(arcsec)} & 
		\colhead{(kpc)} & 
		\colhead{(AB mag.arcsec$^{-2}$)}\\
	}
	\startdata
\cutinhead{Lynx Group 1 ($\langle z \rangle = 1.262$)}
518 & 08 49 03.52 & +44 53 21.62 & 4.7 & 0.24 & 1.99 & 19.1 \\
 &  &  & 4 & 0.21 & 1.72 & 18.8 \\
1339 & 08 49 08.32 & +44 53 48.32 & 4.5 & 0.54 & 4.48 & 19.6 \\
 &  &  & 4 & 0.46 & 3.87 & 19.2 \\
1024 & 08 49 09.00 & +44 52 44.08 & 3.8 & 0.08 & 0.71 & 16.4 \\
 &  &  & 4 & 0.09 & 0.72 & 16.4 \\
825 & 08 49 11.24 & +44 51 29.19 & 4.5 & 0.22 & 1.87 & 17.3 \\
 &  &  & 4 & 0.20 & 1.70 & 17.1 \\
1249 & 08 49 12.27 & +44 52 13.05 & ... & ... & ... & ... \\
1085 & 08 49 13.69 & +44 51 18.82 & 2.7 & 0.23 & 1.94 & 18.0 \\
 &  &  & 4 & 0.31 & 2.62 & 18.6 \\
\cutinhead{Lynx Group 2 ($\langle z \rangle = 1.260$)}
1636 & 08 49 00.92 & +44 58 49.15 & 3.3 & 0.31 & 2.58 & 18.8 \\
 &  &  & 4 & 0.37 & 3.11 & 19.2 \\
1383 & 08 49 03.99 & +44 57 23.37 & 3.2 & 0.19 & 1.60 & 18.2 \\
 &  &  & 4 & 0.23 & 1.91 & 18.5 \\
2000 & 08 49 07.15 & +44 57 52.04 & 5.6 & 1.06 & 8.83 & 21.1 \\
 &  &  & 4 & 0.58 & 4.83 & 19.8 \\
\cutinhead{Lynx Group 3 ($\langle z \rangle = 1.263$)}
137 & 08 48 53.26 & +44 44 22.39 & ... & ... & ... & ... \\
542 & 08 48 55.14 & +44 44 58.83 & ... & ... & ... & ... \\
1135 & 08 48 56.28 & +44 46 45.62 & 5.9 & 1.13 & 9.45 & 21.5 \\
 &  &  & 4 & 0.62 & 5.15 & 20.2 \\
889 & 08 48 56.63 & +44 45 39.90 & 1.5 & 0.15 & 1.25 & 18.5 \\
 &  &  & 4 & 0.25 & 2.08 & 19.6 \\
1431 & 08 48 57.31 & +44 47 08.01 & 4.7 & 0.23 & 1.95 & 18.9 \\
 &  &  & 4 & 0.20 & 1.67 & 18.6 \\
1064 & 08 48 57.79 & +44 45 57.51 & 1.9 & 0.13 & 1.05 & 17.8 \\
 &  &  & 4 & 0.19 & 1.55 & 18.7 \\
1136 & 08 48 57.96 & +44 46 04.53 & 5.9 & 0.48 & 4.00 & 20.0 \\
 &  &  & 4 & 0.30 & 2.53 & 19.0 \\
1775 & 08 49 01.62 & +44 46 28.23 & 6.6 & 1.00 & 8.32 & 21.8 \\
 &  &  & 4 & 0.43 & 3.62 & 20.0 \\
1731 & 08 49 04.43 & +44 45 08.65 & 5.7 & 0.67 & 5.60 & 20.1 \\
 &  &  & 4 & 0.41 & 3.40 & 19.1 \\
\enddata
\tablecomments{$R_e$ denotes the circularized effective radius. Uncertainties on $R_e$ are of 20\%, uncertainties on $n$ are of 20\% and uncertainties on $\langle \mu^B \rangle_e$ are of 0.4 magnitudes. We did not report parameters considered as non robust.}
\end{deluxetable}

\begin{deluxetable}{l c c c c c c c}
	\setlength{\tabcolsep}{0.15in}
	\tablecolumns{7}
	\tablecaption{CDF-S ETG structural parameters and surface brightness. \label{tab:cat_cdfs}}
	\tablehead{
		\colhead{ID $\; ^a$} & 
		\colhead{$z_{spec}$} & 
		\colhead{R.A.} &
		\colhead{DEC.} &
		\colhead{$n$} &
		\colhead{$R_e$}	& 
		\colhead{$R_e$}	& 
		\colhead{$\langle \mu^B \rangle_e$} \\
		\colhead{} & 
		\colhead{} & 
		\colhead{(J2000)} & 
		\colhead{(J2000)} & 
		\colhead{} & 
		\colhead{(arcsec)} & 
		\colhead{(kpc)} & 
		\colhead{(AB mag.arcsec$^{-2}$)}\\
	}
	\startdata

3680 & 1.119 & 03 32 20.28 & -27 52 33.01 & 4.1 & 0.07 & 0.61 & 15.1 \\
 &  &  &  & 4 & 0.07 & 0.61 & 15.1 \\
10069 & 1.119 & 03 32 19.36 & -27 47 16.24 & 2.9 & 0.13 & 1.04 & 17.7 \\
 &  &  &  & 4 & 0.16 & 1.27 & 18.1 \\
7237 & 1.123 & 03 32 45.14 & -27 49 39.95 & 6.5 & 0.19 & 1.52 & 16.8 \\
 &  &  &  & 4 & 0.12 & 0.99 & 15.9 \\
3000 & 1.125 & 03 32 23.60 & -27 53 06.35 & 6.5 & 0.31 & 2.57 & 18.4 \\
 &  &  &  & 4 & 0.18 & 1.49 & 17.2 \\
7567 & 1.158 & 03 32 23.28 & -27 49 26.07 & 5.0 & 0.15 & 1.25 & 17.3 \\
 &  &  &  & 4 & 0.13 & 1.05 & 16.9 \\
10717 & 1.173 & 03 32 30.83 & -27 46 48.56 & 3.2 & 0.17 & 1.39 & 17.8 \\
 &  &  &  & 4 & 0.21 & 1.71 & 18.3 \\
14747 & 1.178 & 03 32 39.17 & -27 43 29.02 & 3.6 & 0.34 & 2.85 & 19.7 \\
 &  &  &  & 4 & 0.39 & 3.24 & 19.9 \\
9066 & 1.188 & 03 32 33.06 & -27 48 07.54 & 4.3 & 0.34 & 2.79 & 19.4 \\
 &  &  &  & 4 & 0.31 & 2.56 & 19.2 \\
4176 & 1.189 & 03 32 24.98 & -27 52 08.63 & 4.3 & 0.22 & 1.85 & 18.8 \\
 &  &  &  & 4 & 0.21 & 1.74 & 18.6 \\
14953 & 1.215 & 03 32 25.98 & -27 43 18.93 & 5.3 & 0.20 & 1.67 & 17.8 \\
 &  &  &  & 4 & 0.16 & 1.30 & 17.3 \\
11062 & 1.220 & 03 32 46.34 & -27 46 32.00 & 4.4 & 0.50 & 4.18 & 20.7 \\
 &  &  &  & 4 & 0.45 & 3.72 & 20.4 \\
15093 & 1.222 & 03 32 35.63 & -27 43 10.14 & 6.3 & 0.42 & 3.45 & 18.6 \\
 &  &  &  & 4 & 0.24 & 1.96 & 17.4 \\
12264 & 1.222 & 03 32 26.29 & -27 45 36.19 & 1.3 & 0.07 & 0.55 & 14.6 \\
 &  &  &  & 4 & 0.09 & 0.74 & 15.3 \\
12000 & 1.222 & 03 32 26.26 & -27 45 50.71 & 5.9 & 0.31 & 2.59 & 18.6 \\
 &  &  &  & 4 & 0.21 & 1.73 & 17.7 \\
9702 & 1.223 & 03 32 35.79 & -27 47 34.76 & 5.0 & 0.26 & 2.18 & 19.2 \\
 &  &  &  & 4 & 0.21 & 1.74 & 18.8 \\
4981 & 1.253 & 03 32 44.26 & -27 51 26.75 & 6.9 & 0.30 & 2.52 & 18.5 \\
 &  &  &  & 4 & 0.17 & 1.39 & 17.2 \\
288 & 1.264 & 03 32 25.40 & -27 56 09.88 & 5.1 & 0.06 & 0.51 & 16.0 \\
 &  &  &  & 4 & 0.06 & 0.47 & 15.8 \\
10650 & 1.277 & 03 32 08.37 & -27 46 51.21 & 8.2 & 0.22 & 1.82 & 17.5 \\
 &  &  &  & 4 & 0.11 & 0.92 & 16.1 \\
6791 & 1.297 & 03 32 50.19 & -27 50 01.04 & ... & ... & ... & ... \\
9369 & 1.297 & 03 32 16.02 & -27 47 50.00 & 7.7 & 0.62 & 5.21 & 20.1 \\
 &  &  &  & 4 & 0.24 & 2.00 & 18.0 \\
10231 & 1.317 & 03 32 39.63 & -27 47 09.12 & 5.4 & 0.92 & 7.74 & 20.8 \\
 &  &  &  & 4 & 0.58 & 4.86 & 19.8 \\
17506 & 1.328 & 03 32 20.08 & -27 41 06.75 & 6.1 & 0.42 & 3.50 & 18.6 \\
 &  &  &  & 4 & 0.25 & 2.07 & 17.5 \\
1857 & 1.345 & 03 32 38.37 & -27 54 08.83 & 5.2 & 0.42 & 3.56 & 19.4 \\
 &  &  &  & 4 & 0.31 & 2.61 & 18.8 \\
969 & 1.346 & 03 32 35.99 & -27 55 09.49 & 7.0 & 0.55 & 4.61 & 20.2 \\
 &  &  &  & 4 & 0.26 & 2.15 & 18.5 \\
10041 & 1.356 & 03 32 25.04 & -27 47 18.20 & 0.6 & 0.13 & 1.12 & 17.6 \\
 &  &  &  & 4 & 0.29 & 2.41 & 19.2 \\
12505 & 1.374 & 03 32 06.81 & -27 45 24.35 & 3.8 & 0.14 & 1.15 & 16.9 \\
 &  &  &  & 4 & 0.14 & 1.18 & 17.0 \\
8938 & 1.382 & 03 32 33.98 & -27 48 14.69 & 1.4 & 0.08 & 0.70 & 16.6 \\
 &  &  &  & 4 & 0.12 & 0.99 & 17.4 \\
\enddata
\tablecomments{$\; ^a$ ID refers to the GOODS-MUSIC v2 catalogue of \citet{santini09}. $R_e$ denotes the circularized effective radius. Uncertainties on $R_e$ are of 20\%, uncertainties on $n$ are of 20\% and uncertainties on $\langle \mu^B \rangle_e$ are of 0.4 magnitudes. We did not report parameters considered as non robust.}
\end{deluxetable}

\newpage


\section{Appendix B: Further tests and explanations \label{sec:supfig}}

This appendix shows some of the tests explained in the text.

\begin{figure*}
	\includegraphics[width=\linewidth]{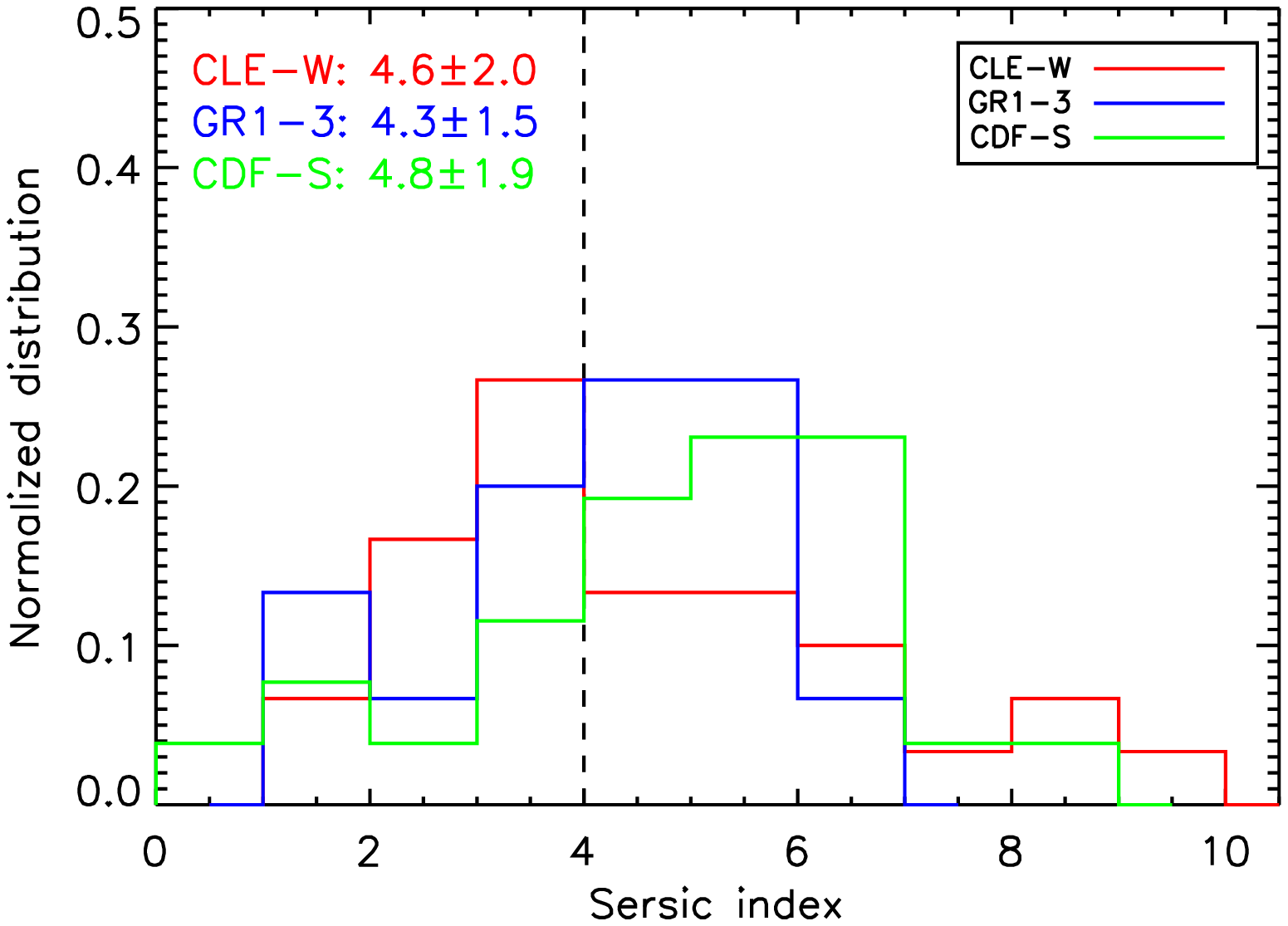}
	\caption{
S\'ersic index distributions for our sample:
red histogram is for cluster ETGs, blue histogram is for group ETGs and green histogram is for field ETGs.
The dashed line represents a S\'ersic index of 4, corresponding to a de Vaucouleurs profile.
We also report the mean and standard deviation for the three distributions.
}
	\label{fig:nser_hist}
\end{figure*}

Figure \ref{fig:nser_hist} show the S\'ersic index distribution for our sample \citep[see also][]{mei11}.

\begin{figure*}
	\includegraphics[width=\linewidth]{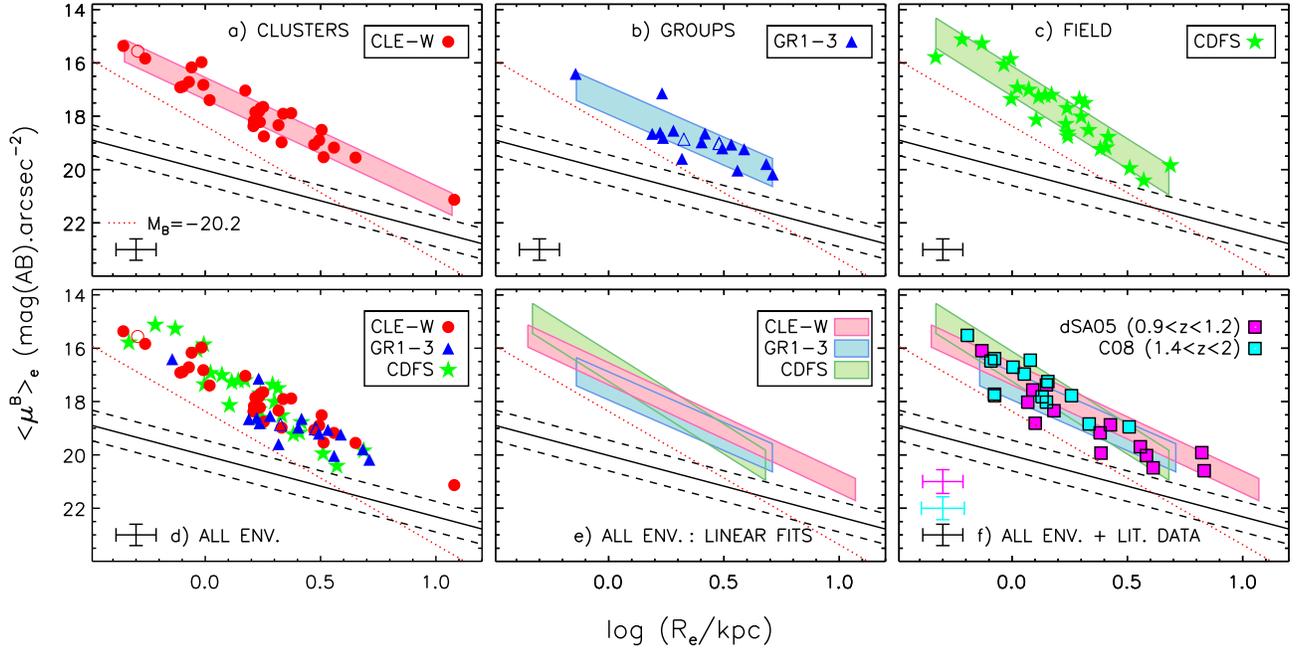}
	\caption{
Same as Figure \ref{fig:kr}, but with our sizes estimated with a de Vaucouleurs profile.
Moreover, we include here in the local KR of \citet{jorgensen95} the three largest galaxies that we excluded in Figure \ref{fig:kr}.
We observe that none of our conclusions of Section \ref{sec:kr} is affected if we use de Vaucouleurs sizes.
}
	\label{fig:kr_vauc}
\end{figure*}

Figure \ref{fig:kr_vauc} compares our sizes derived with a de Vaucouleurs profile to the local KR, derived from the 31 ETGs measurements of \citet{jorgensen95}.
We show that our results from Section \ref{sec:kr} do not change.

\begin{figure*}
	\includegraphics[width=\linewidth]{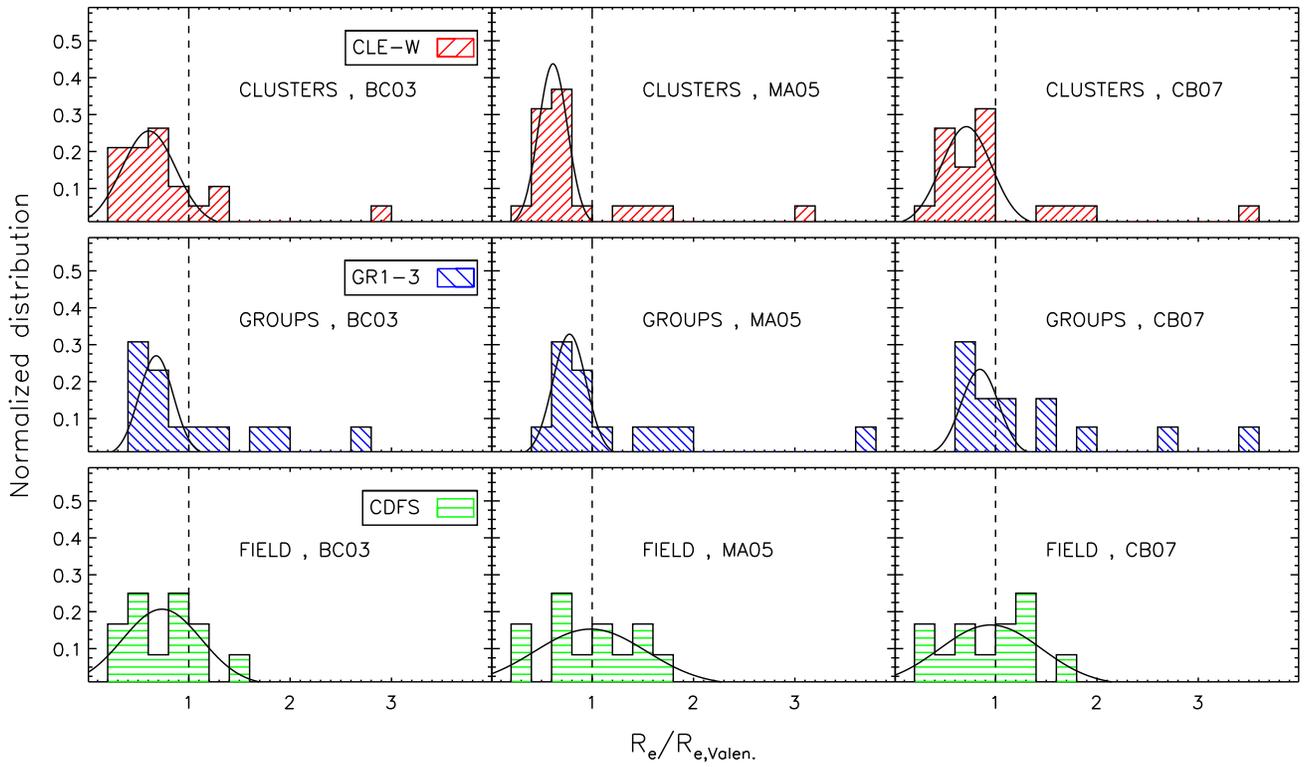}
	\caption{
Size ratio $R_e/R_{e,Valen}$ normalized distributions derived with the three stellar population models (BC03/MA05/CB07), split by environments:
same figure as Figure \ref{fig:msr_hist}, but we here remove ETGs known to have emission lines (7 ETGs in clusters, 4 ETGs in groups and 14 ETGs in the field). Our results do not change.
}
	\label{fig:msr_hist_noem}
\end{figure*}

Figure \ref{fig:msr_hist_noem} is similar to the figure \ref{fig:msr_hist}, but we removed from the sample the ETGs known to have emission lines and as a consequence of it, larger sizes (see Section 5.1). It shows that even if we do remove these galaxies our results do not change.

\begin{figure*}
	\includegraphics[width=\linewidth]{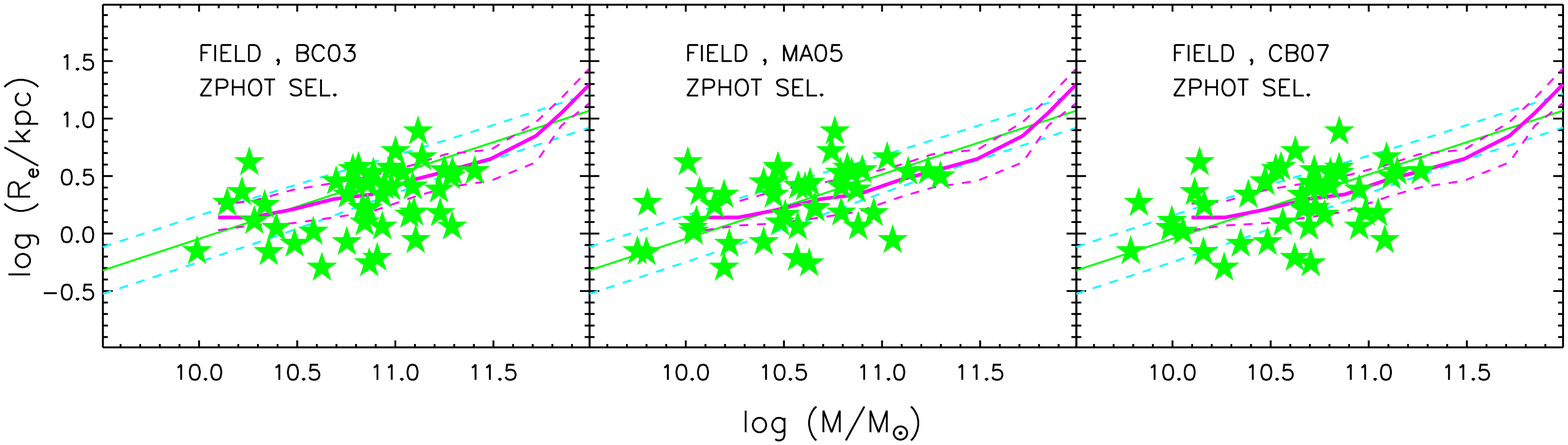}
	\includegraphics[width=\linewidth]{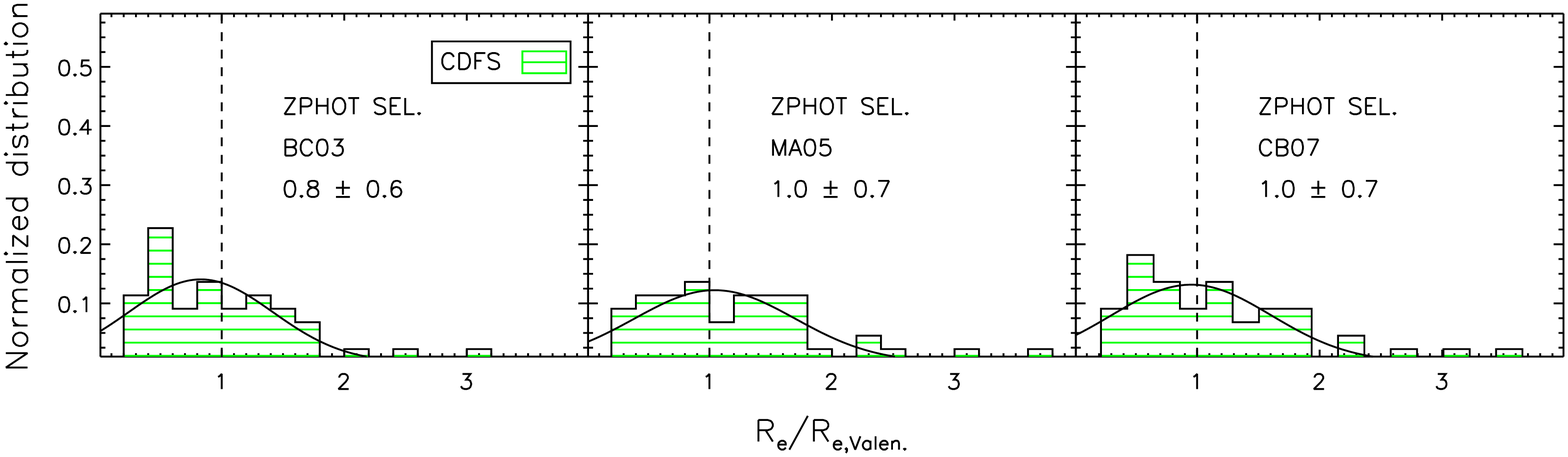}
	\caption{
Results when considering a CDF-S sample including ETGs with photometric redshift.
\textit{Upper panels}: MSR, symbols are as in Figure~\ref{fig:msr}.
\textit{Lower panels}: size ratio $R_e/R_{e,Valen}$ normalized distributions derived with the three stellar population models (BC03/MA05/CB07), same as Figure \ref{fig:msr_hist}.
This sample shows a MSR similar to the local, as the spectroscopically selected sample.
This test shows that our spectroscopically selected sample is not biased with respect to a photometric redshift selection.
}
	\label{fig:msr_cdfs_test}
\end{figure*}

Figure \ref{fig:msr_cdfs_test} shows how the MSR in the field would change if we add to our CDF-S spectroscopic sample ETGs that are selected using photometric redshifts.
To perform this test, we add to our spectroscopic sample a sample of ETGs selected to have with $1 \le z_{phot} \le 1.5$ and no reliable $z_{spec}$ \citep{santini09}, in the same magnitude and color range that our spectroscopic sample
We thus include any  possible low-mass/passive ETGs which may be absent from our CDF-S sample.
We remark that this test sample might be contaminated by outliers.
For the new ETGs included in the sample, we derive masses and sizes with the same procedure used in this paper.
Figure \ref{fig:msr_cdfs_test}  shows the MSR for this test sample (upper panels) and the size ratio $R_e/R_{e,Valen}$ normalized distributions (lower panels).
The distribution of the size ratios $R_e/R_{e,Valen}$ is similar to the one we find in Figure \ref{fig:msr_hist}.
Our results do not change: the field MSR has still a distribution similar to the local relation.
In the main body of the paper, we leave results obtained using only field spectroscopically confirmed members, since we expect that the photometric redshift selected sample in the field might present a much higher contamination than in clusters and groups.

\begin{thebibliography}{89}
\expandafter\ifx\csname natexlab\endcsname\relax\def\natexlab#1{#1}\fi

\bibitem[{{Bernardi} {et~al.}(2010){Bernardi}, {Shankar}, {Hyde}, {Mei},
  {Marulli}, \& {Sheth}}]{bernardi10}
{Bernardi}, M., {Shankar}, F., {Hyde}, J.~B., {et~al.} 2010, \mnras, 404, 2087

\bibitem[{{Bernardi} {et~al.}(2003){Bernardi}, {Sheth}, {Annis}, {Burles},
  {Eisenstein}, {Finkbeiner}, {Hogg}, {Lupton}, {Schlegel}, {SubbaRao},
  {Bahcall}, {Blakeslee}, {Brinkmann}, {Castander}, {Connolly}, {Csabai},
  {Doi}, {Fukugita}, {Frieman}, {Heckman}, {Hennessy}, {Ivezi{\'c}}, {Knapp},
  {Lamb}, {McKay}, {Munn}, {Nichol}, {Okamura}, {Schneider}, {Thakar}, \&
  {York}}]{bernardi03a}
{Bernardi}, M., {Sheth}, R.~K., {Annis}, J., {et~al.} 2003, \aj, 125, 1817

\bibitem[{{Bertin} \& {Arnouts}(1996)}]{bertin96}
{Bertin}, E. \& {Arnouts}, S. 1996, \aaps, 117, 393

\bibitem[{{Bezanson} {et~al.}(2009){Bezanson}, {van Dokkum}, {Tal},
  {Marchesini}, {Kriek}, {Franx}, \& {Coppi}}]{bezanson09}
{Bezanson}, R., {van Dokkum}, P.~G., {Tal}, T., {et~al.} 2009, \apj, 697, 1290

\bibitem[{{Blanton} {et~al.}(2005){Blanton}, {Eisenstein}, {Hogg}, {Schlegel},
  \& {Brinkmann}}]{blanton05}
{Blanton}, M.~R., {Eisenstein}, D., {Hogg}, D.~W., {Schlegel}, D.~J., \&
  {Brinkmann}, J. 2005, \apj, 629, 143

\bibitem[{{Bournaud} {et~al.}(2011){Bournaud}, {Chapon}, {Teyssier}, {Powell},
  {Elmegreen}, {Elmegreen}, {Duc}, {Contini}, {Epinat}, \&
  {Shapiro}}]{bournaud11}
{Bournaud}, F., {Chapon}, D., {Teyssier}, R., {et~al.} 2011, \apj, 730, 4

\bibitem[{{Bruzual} \& {Charlot}(2003)}]{bruzual03}
{Bruzual}, G. \& {Charlot}, S. 2003, \mnras, 344, 1000

\bibitem[{{Buitrago} {et~al.}(2008){Buitrago}, {Trujillo}, {Conselice},
  {Bouwens}, {Dickinson}, \& {Yan}}]{buitrago08}
{Buitrago}, F., {Trujillo}, I., {Conselice}, C.~J., {et~al.} 2008, \apjl, 687,
  L61

\bibitem[{{Caon} {et~al.}(1993){Caon}, {Capaccioli}, \& {D'Onofrio}}]{caon93}
{Caon}, N., {Capaccioli}, M., \& {D'Onofrio}, M. 1993, \mnras, 265, 1013

\bibitem[{Cassata {et~al.}(2011)Cassata, Giavalisco, Guo, Renzini, Ferguson,
  Koekemoer, Salimbeni, Scarlata, Grogin, Conselice, Dahlen, Lotz, Dickinson,
  \& Lin}]{cassata11}
Cassata, P., Giavalisco, M., Guo, Y., {et~al.} 2011

\bibitem[{{Cimatti} {et~al.}(2008){Cimatti}, {Cassata}, {Pozzetti}, {Kurk},
  {Mignoli}, {Renzini}, {Daddi}, {Bolzonella}, {Brusa}, {Rodighiero},
  {Dickinson}, {Franceschini}, {Zamorani}, {Berta}, {Rosati}, \&
  {Halliday}}]{cimatti08}
{Cimatti}, A., {Cassata}, P., {Pozzetti}, L., {et~al.} 2008, \aap, 482, 21

\bibitem[{{Conroy} \& {Gunn}(2010)}]{conroy10}
{Conroy}, C. \& {Gunn}, J.~E. 2010, \apj, 712, 833

\bibitem[{{Daddi} {et~al.}(2005){Daddi}, {Renzini}, {Pirzkal}, {Cimatti},
  {Malhotra}, {Stiavelli}, {Xu}, {Pasquali}, {Rhoads}, {Brusa}, {di Serego
  Alighieri}, {Ferguson}, {Koekemoer}, {Moustakas}, {Panagia}, \&
  {Windhorst}}]{daddi05}
{Daddi}, E., {Renzini}, A., {Pirzkal}, N., {et~al.} 2005, \apj, 626, 680

\bibitem[{{Damjanov} {et~al.}(2009){Damjanov}, {McCarthy}, {Abraham},
  {Glazebrook}, {Yan}, {Mentuch}, {Le Borgne}, {Savaglio}, {Crampton},
  {Murowinski}, {Juneau}, {Carlberg}, {J{\o}rgensen}, {Roth}, {Chen}, \&
  {Marzke}}]{damjanov09}
{Damjanov}, I., {McCarthy}, P.~J., {Abraham}, R.~G., {et~al.} 2009, \apj, 695,
  101

\bibitem[{{de Vaucouleurs}(1948)}]{de-vaucouleurs48}
{de Vaucouleurs}, G. 1948, Annales d'Astrophysique, 11, 247

\bibitem[{{di Serego Alighieri} {et~al.}(2005){di Serego Alighieri}, {Vernet},
  {Cimatti}, {Lanzoni}, {Cassata}, {Ciotti}, {Daddi}, {Mignoli}, {Pignatelli},
  {Pozzetti}, {Renzini}, {Rettura}, \& {Zamorani}}]{di-serego-alighieri05}
{di Serego Alighieri}, S., {Vernet}, J., {Cimatti}, A., {et~al.} 2005, \aap,
  442, 125

\bibitem[{{Djorgovski} \& {Davis}(1987)}]{djorgovski87}
{Djorgovski}, S. \& {Davis}, M. 1987, \apj, 313, 59

\bibitem[{{D'Onofrio} {et~al.}(2008){D'Onofrio}, {Fasano}, {Varela}, {Bettoni},
  {Moles}, {Kj{\ae}rgaard}, {Pignatelli}, {Poggianti}, {Dressler}, {Cava},
  {Fritz}, {Couch}, \& {Omizzolo}}]{donofrio08}
{D'Onofrio}, M., {Fasano}, G., {Varela}, J., {et~al.} 2008, \apj, 685, 875

\bibitem[{{Dressler} {et~al.}(1987){Dressler}, {Lynden-Bell}, {Burstein},
  {Davies}, {Faber}, {Terlevich}, \& {Wegner}}]{dressler87}
{Dressler}, A., {Lynden-Bell}, D., {Burstein}, D., {et~al.} 1987, \apj, 313, 42

\bibitem[{{Fan} {et~al.}(2008){Fan}, {Lapi}, {De Zotti}, \& {Danese}}]{fan08}
{Fan}, L., {Lapi}, A., {De Zotti}, G., \& {Danese}, L. 2008, \apjl, 689, L101

\bibitem[{{Franx} {et~al.}(2008){Franx}, {van Dokkum}, {Schreiber}, {Wuyts},
  {Labb{\'e}}, \& {Toft}}]{franx08}
{Franx}, M., {van Dokkum}, P.~G., {Schreiber}, N.~M.~F., {et~al.} 2008, \apj,
  688, 770

\bibitem[{{Geller} \& {Huchra}(1983)}]{geller83}
{Geller}, M.~J. \& {Huchra}, J.~P. 1983, \apjs, 52, 61

\bibitem[{{Giavalisco} {et~al.}(2004){Giavalisco}, {Ferguson}, {Koekemoer},
  {Dickinson}, {Alexander}, {Bauer}, {Bergeron}, {Biagetti}, {Brandt},
  {Casertano}, {Cesarsky}, {Chatzichristou}, {Conselice}, {Cristiani}, {Da
  Costa}, {Dahlen}, {de Mello}, {Eisenhardt}, {Erben}, {Fall}, {Fassnacht},
  {Fosbury}, {Fruchter}, {Gardner}, {Grogin}, {Hook}, {Hornschemeier}, {Idzi},
  {Jogee}, {Kretchmer}, {Laidler}, {Lee}, {Livio}, {Lucas}, {Madau},
  {Mobasher}, {Moustakas}, {Nonino}, {Padovani}, {Papovich}, {Park},
  {Ravindranath}, {Renzini}, {Richardson}, {Riess}, {Rosati}, {Schirmer},
  {Schreier}, {Somerville}, {Spinrad}, {Stern}, {Stiavelli}, {Strolger},
  {Urry}, {Vandame}, {Williams}, \& {Wolf}}]{giavalisco04}
{Giavalisco}, M., {Ferguson}, H.~C., {Koekemoer}, A.~M., {et~al.} 2004, \apjl,
  600, L93

\bibitem[{{Granato} {et~al.}(2006){Granato}, {Silva}, {Lapi}, {Shankar}, {De
  Zotti}, \& {Danese}}]{granato06}
{Granato}, G.~L., {Silva}, L., {Lapi}, A., {et~al.} 2006, \mnras, 368, L72

\bibitem[{{H{\"a}ussler} {et~al.}(2007){H{\"a}ussler}, {McIntosh}, {Barden},
  {Bell}, {Rix}, {Borch}, {Beckwith}, {Caldwell}, {Heymans}, {Jahnke}, {Jogee},
  {Koposov}, {Meisenheimer}, {S{\'a}nchez}, {Somerville}, {Wisotzki}, \&
  {Wolf}}]{haussler07}
{H{\"a}ussler}, B., {McIntosh}, D.~H., {Barden}, M., {et~al.} 2007, \apjs, 172,
  615

\bibitem[{{Holden} {et~al.}(2005){Holden}, {Blakeslee}, {Postman},
  {Illingworth}, {Demarco}, {Franx}, {Rosati}, {Bouwens}, {Martel}, {Ford},
  {Clampin}, {Hartig}, {Ben{\'{\i}}tez}, {Cross}, {Homeier}, {Lidman},
  {Menanteau}, {Zirm}, {Ardila}, {Bartko}, {Bradley}, {Broadhurst}, {Brown},
  {Burrows}, {Cheng}, {Feldman}, {Golimowski}, {Goto}, {Gronwall}, {Infante},
  {Kimble}, {Krist}, {Lesser}, {Magee}, {Mei}, {Meurer}, {Miley}, {Motta},
  {Sirianni}, {Sparks}, {Tran}, {Tsvetanov}, {White}, \& {Zheng}}]{holden05}
{Holden}, B.~P., {Blakeslee}, J.~P., {Postman}, M., {et~al.} 2005, \apj, 626,
  809

\bibitem[{{Hopkins} {et~al.}(2009{\natexlab{a}}){Hopkins}, {Bundy}, {Murray},
  {Quataert}, {Lauer}, \& {Ma}}]{hopkins09}
{Hopkins}, P.~F., {Bundy}, K., {Murray}, N., {et~al.} 2009{\natexlab{a}},
  \mnras, 398, 898

\bibitem[{{Hopkins} {et~al.}(2009{\natexlab{b}}){Hopkins}, {Hernquist}, {Cox},
  {Keres}, \& {Wuyts}}]{hopkins09b}
{Hopkins}, P.~F., {Hernquist}, L., {Cox}, T.~J., {Keres}, D., \& {Wuyts}, S.
  2009{\natexlab{b}}, \apj, 691, 1424

\bibitem[{{Hopkins} {et~al.}(2010){Hopkins}, {Bundy}, {Hernquist}, {Wuyts}, \&
  {Cox}}]{hopkins10}
{Hopkins}, P.~F., {Bundy}, K., {Hernquist}, L., {Wuyts}, S., \& {Cox}, T.~J.
  2010, \mnras, 401, 1099

\bibitem[{{Jorgensen} {et~al.}(1995){Jorgensen}, {Franx}, \&
  {Kjaergaard}}]{jorgensen95}
{Jorgensen}, I., {Franx}, M., \& {Kjaergaard}, P. 1995, \mnras, 273, 1097

\bibitem[{{Kaviraj} {et~al.}(2009){Kaviraj}, {Devriendt}, {Ferreras}, {Yi}, \&
  {Silk}}]{kaviraj09}
{Kaviraj}, S., {Devriendt}, J.~E.~G., {Ferreras}, I., {Yi}, S.~K., \& {Silk},
  J. 2009, \aap, 503, 445

\bibitem[{{Khochfar} \& {Silk}(2006)}]{khochfar06a}
{Khochfar}, S. \& {Silk}, J. 2006, \mnras, 370, 902

\bibitem[{{Kormendy}(1977)}]{kormendy77}
{Kormendy}, J. 1977, \apj, 218, 333

\bibitem[{{Kron}(1980)}]{kron80}
{Kron}, R.~G. 1980, \apjs, 43, 305

\bibitem[{{Kroupa}(2001)}]{kroupa01}
{Kroupa}, P. 2001, \mnras, 322, 231

\bibitem[{{La Barbera} {et~al.}(2003){La Barbera}, {Busarello}, {Merluzzi},
  {Massarotti}, \& {Capaccioli}}]{la-barbera03}
{La Barbera}, F., {Busarello}, G., {Merluzzi}, P., {Massarotti}, M., \&
  {Capaccioli}, M. 2003, \apj, 595, 127

\bibitem[{{Longhetti} {et~al.}(2007){Longhetti}, {Saracco}, {Severgnini},
  {Della Ceca}, {Mannucci}, {Bender}, {Drory}, {Feulner}, \&
  {Hopp}}]{longhetti07}
{Longhetti}, M., {Saracco}, P., {Severgnini}, P., {et~al.} 2007, \mnras, 374,
  614

\bibitem[{{Maltby} {et~al.}(2010){Maltby}, {Arag{\'o}n-Salamanca}, {Gray},
  {Barden}, {H{\"a}u{\ss}ler}, {Wolf}, {Peng}, {Jahnke}, {McIntosh},
  {B{\"o}hm}, \& {van Kampen}}]{maltby10}
{Maltby}, D.~T., {Arag{\'o}n-Salamanca}, A., {Gray}, M.~E., {et~al.} 2010,
  \mnras, 402, 282

\bibitem[{{Mancini} {et~al.}(2010){Mancini}, {Daddi}, {Renzini}, {Salmi},
  {McCracken}, {Cimatti}, {Onodera}, {Salvato}, {Koekemoer}, {Aussel},
  {Floc'h}, {Willott}, \& {Capak}}]{mancini10}
{Mancini}, C., {Daddi}, E., {Renzini}, A., {et~al.} 2010, \mnras, 401, 933

\bibitem[{{Maraston}(2005)}]{maraston05}
{Maraston}, C. 2005, \mnras, 362, 799

\bibitem[{{Maraston} {et~al.}(2006){Maraston}, {Daddi}, {Renzini}, {Cimatti},
  {Dickinson}, {Papovich}, {Pasquali}, \& {Pirzkal}}]{maraston06}
{Maraston}, C., {Daddi}, E., {Renzini}, A., {et~al.} 2006, \apj, 652, 85

\bibitem[{{McGrath} {et~al.}(2008){McGrath}, {Stockton}, {Canalizo}, {Iye}, \&
  {Maihara}}]{mcgrath08}
{McGrath}, E.~J., {Stockton}, A., {Canalizo}, G., {Iye}, M., \& {Maihara}, T.
  2008, \apj, 682, 303

\bibitem[{{Mei} {et~al.}(2011){Mei}}]{mei11}
{Mei}, S., {et~al.} 2011, \apj, submitted

\bibitem[{{Mei} {et~al.}(2009){Mei}, {Holden}, {Blakeslee}, {Ford}, {Franx},
  {Homeier}, {Illingworth}, {Jee}, {Overzier}, {Postman}, {Rosati}, {Van der
  Wel}, \& {Bartlett}}]{mei09}
{Mei}, S., {Holden}, B.~P., {Blakeslee}, J.~P., {et~al.} 2009, \apj, 690, 42

\bibitem[{{Mei} {et~al.}(2006{\natexlab{a}}){Mei}, {Blakeslee}, {Stanford},
  {Holden}, {Rosati}, {Strazzullo}, {Homeier}, {Postman}, {Franx}, {Rettura},
  {Ford}, {Illingworth}, {Ettori}, {Bouwens}, {Demarco}, {Martel}, {Clampin},
  {Hartig}, {Eisenhardt}, {Ardila}, {Bartko}, {Ben{\'{\i}}tez}, {Bradley},
  {Broadhurst}, {Brown}, {Burrows}, {Cheng}, {Cross}, {Feldman}, {Golimowski},
  {Goto}, {Gronwall}, {Infante}, {Kimble}, {Krist}, {Lesser}, {Menanteau},
  {Meurer}, {Miley}, {Motta}, {Sirianni}, {Sparks}, {Tran}, {Tsvetanov},
  {White}, \& {Zheng}}]{mei06a}
{Mei}, S., {Blakeslee}, J.~P., {Stanford}, S.~A., {et~al.} 2006{\natexlab{a}},
  \apj, 639, 81

\bibitem[{{Mei} {et~al.}(2006{\natexlab{b}}){Mei}, {Holden}, {Blakeslee},
  {Rosati}, {Postman}, {Jee}, {Rettura}, {Sirianni}, {Demarco}, {Ford},
  {Franx}, {Homeier}, \& {Illingworth}}]{mei06}
{Mei}, S., {Holden}, B.~P., {Blakeslee}, J.~P., {et~al.} 2006{\natexlab{b}},
  \apj, 644, 759
  
\bibitem[{{Naab} {et~al.}(2009){Naab}, {Johansson}, \& {Ostriker}}]{naab09}
{Naab}, T., {Johansson}, P.~H., \& {Ostriker}, J.~P. 2009, \apjl, 699, L178

\bibitem[{{Nakata} {et~al.}(2005){Nakata}, {Kodama}, {Shimasaku}, {Doi},
  {Furusawa}, {Hamabe}, {Kimura}, {Komiyama}, {Miyazaki}, {Okamura}, {Ouchi},
  {Sekiguchi}, {Ueda}, {Yagi}, \& {Yasuda}}]{nakata05}
{Nakata}, F., {Kodama}, T., {Shimasaku}, K., {et~al.} 2005, \mnras, 357, 1357

\bibitem[{{Newman} {et~al.}(2010){Newman}, {Ellis}, {Treu}, \&
  {Bundy}}]{newman10}
{Newman}, A.~B., {Ellis}, R.~S., {Treu}, T., \& {Bundy}, K. 2010, \apjl, 717,
  L103

\bibitem[{{Nonino} {et~al.}(2009){Nonino}, {Dickinson}, {Rosati}, {Grazian},
  {Reddy}, {Cristiani}, {Giavalisco}, {Kuntschner}, {Vanzella}, {Daddi},
  {Fosbury}, \& {Cesarsky}}]{nonino09}
{Nonino}, M., {Dickinson}, M., {Rosati}, P., {et~al.} 2009, \apjs, 183, 244

\bibitem[{{Onodera} {et~al.}(2010){Onodera}, {Daddi}, {Gobat}, {Cappellari},
  {Arimoto}, {Renzini}, {Yamada}, {McCracken}, {Mancini}, {Capak}, {Carollo},
  {Cimatti}, {Giavalisco}, {Ilbert}, {Kong}, {Lilly}, {Motohara}, {Ohta},
  {Sanders}, {Scoville}, {Tamura}, \& {Taniguchi}}]{onodera10}
{Onodera}, M., {Daddi}, E., {Gobat}, R., {et~al.} 2010, \apjl, 715, L6

\bibitem[{{Peng} {et~al.}(2002){Peng}, {Ho}, {Impey}, \& {Rix}}]{peng02}
{Peng}, C.~Y., {Ho}, L.~C., {Impey}, C.~D., \& {Rix}, H. 2002, \aj, 124, 266

\bibitem[{{Poggianti} {et~al.}(2006){Poggianti}, {von der Linden}, {De Lucia},
  {Desai}, {Simard}, {Halliday}, {Arag{\'o}n-Salamanca}, {Bower}, {Varela},
  {Best}, {Clowe}, {Dalcanton}, {Jablonka}, {Milvang-Jensen}, {Pello},
  {Rudnick}, {Saglia}, {White}, \& {Zaritsky}}]{poggianti06}
{Poggianti}, B.~M., {von der Linden}, A., {De Lucia}, G., {et~al.} 2006, \apj,
  642, 188

\bibitem[{{Postman} {et~al.}(2005){Postman}, {Franx}, {Cross}, {Holden},
  {Ford}, {Illingworth}, {Goto}, {Demarco}, {Rosati}, {Blakeslee}, {Tran},
  {Ben{\'{\i}}tez}, {Clampin}, {Hartig}, {Homeier}, {Ardila}, {Bartko},
  {Bouwens}, {Bradley}, {Broadhurst}, {Brown}, {Burrows}, {Cheng}, {Feldman},
  {Golimowski}, {Gronwall}, {Infante}, {Kimble}, {Krist}, {Lesser}, {Martel},
  {Mei}, {Menanteau}, {Meurer}, {Miley}, {Motta}, {Sirianni}, {Sparks}, {Tran},
  {Tsvetanov}, {White}, \& {Zheng}}]{postman05}
{Postman}, M., {Franx}, M., {Cross}, N.~J.~G., {et~al.} 2005, \apj, 623, 721

\bibitem[{{Raichoor} {et~al.}(2011){Raichoor}, {Mei}, {Nakata}, {Stanford},
  {Holden}, {Rettura}, {Huertas-Company}, {Postman}, {Rosati}, {Blakeslee},
  {Demarco}, {Eisenhardt}, {Illingworth}, {Jee}, {Kodama}, {Tanaka}, \&
  {White}}]{raichoor11}
{Raichoor}, A., {Mei}, S., {Nakata}, F., {et~al.} 2011, \apj, 732, 12

\bibitem[{{Rettura} {et~al.}(2010){Rettura}, {Rosati}, {Nonino}, {Fosbury},
  {Gobat}, {Menci}, {Strazzullo}, {Mei}, {Demarco}, \& {Ford}}]{rettura10}
{Rettura}, A., {Rosati}, P., {Nonino}, M., {et~al.} 2010, \apj, 709, 512

\bibitem[{{Rettura} {et~al.}(2011){Rettura}, {Mei}, {Stanford}, {Raichoor},
  {Moran}, {Holden}, {Rosati}, {Ellis}, {Nakata}, {Nonino}, {Treu},
  {Blakeslee}, {Demarco}, {Eisenhardt}, {Ford}, {Fosbury}, {Illingworth},
  {Huertas-Company}, {Jee}, {Kodama}, {Postman}, {Tanaka}, \&
  {White}}]{rettura11}
{Rettura}, A., {Mei}, S., {Stanford}, S.~A., {et~al.} 2011, \apj, 732, 94

\bibitem[{{Retzlaff} {et~al.}(2010){Retzlaff}, {Rosati}, {Dickinson},
  {Vandame}, {Rit{\'e}}, {Nonino}, {Cesarsky}, \& {GOODS Team}}]{retzlaff10}
{Retzlaff}, J., {Rosati}, P., {Dickinson}, M., {et~al.} 2010, \aap, 511, A50

\bibitem[{{Rosati} {et~al.}(1999){Rosati}, {Stanford}, {Eisenhardt}, {Elston},
  {Spinrad}, {Stern}, \& {Dey}}]{rosati99}
{Rosati}, P., {Stanford}, S.~A., {Eisenhardt}, P.~R., {et~al.} 1999, \aj, 118,
  76

\bibitem[{{Saglia} {et~al.}(2010){Saglia}, {S{\'a}nchez-Bl{\'a}zquez},
  {Bender}, {Simard}, {Desai}, {Arag{\'o}n-Salamanca}, {Milvang-Jensen},
  {Halliday}, {Jablonka}, {Noll}, {Poggianti}, {Clowe}, {De Lucia},
  {Pell{\'o}}, {Rudnick}, {Valentinuzzi}, {White}, \& {Zaritsky}}]{saglia10}
{Saglia}, R.~P., {S{\'a}nchez-Bl{\'a}zquez}, P., {Bender}, R., {et~al.} 2010,
  \aap, 524, A6

\bibitem[{{Salimbeni} {et~al.}(2009){Salimbeni}, {Castellano}, {Pentericci},
  {Trevese}, {Fiore}, {Grazian}, {Fontana}, {Giallongo}, {Boutsia},
  {Cristiani}, {de Santis}, {Gallozzi}, {Menci}, {Nonino}, {Paris}, {Santini},
  \& {Vanzella}}]{salimbeni09}
{Salimbeni}, S., {Castellano}, M., {Pentericci}, L., {et~al.} 2009, \aap, 501,
  865

\bibitem[{{Salpeter}(1955)}]{salpeter55}
{Salpeter}, E.~E. 1955, \apj, 121, 161

\bibitem[{{Santini} {et~al.}(2009){Santini}, {Fontana}, {Grazian}, {Salimbeni},
  {Fiore}, {Fontanot}, {Boutsia}, {Castellano}, {Cristiani}, {de Santis},
  {Gallozzi}, {Giallongo}, {Menci}, {Nonino}, {Paris}, {Pentericci}, \&
  {Vanzella}}]{santini09}
{Santini}, P., {Fontana}, A., {Grazian}, A., {et~al.} 2009, \aap, 504, 751

\bibitem[{{Saracco} {et~al.}(2009){Saracco}, {Longhetti}, \&
  {Andreon}}]{saracco09}
{Saracco}, P., {Longhetti}, M., \& {Andreon}, S. 2009, \mnras, 392, 718

\bibitem[{{Saracco} {et~al.}(2010){Saracco}, {Longhetti}, \&
  {Gargiulo}}]{saracco10}
{Saracco}, P., {Longhetti}, M., \& {Gargiulo}, A. 2010, \mnras, 408, L21

\bibitem[{{Saracco} {et~al.}(2011){Saracco}, {Longhetti}, \&
  {Gargiulo}}]{saracco11}
{Saracco}, P., {Longhetti}, M., \& {Gargiulo}, A. 2011, \mnras, 187

\bibitem[{{S\'ersic}(1968)}]{sersic68}
{S\'ersic}, J.~L. 1968, {Atlas de galaxias australes (Observatorio Astronomico,
  Cordoba, Argentina)}

\bibitem[{Shankar {et~al.}(2011)Shankar, Marulli, Bernardi, Mei, Meert, \&
  Vikram}]{shankar11}
Shankar, F., Marulli, F., Bernardi, M., {et~al.} 2011, arXiv:1105.6043

\bibitem[{{Shankar} {et~al.}(2010){Shankar}, {Marulli}, {Bernardi},
  {Boylan-Kolchin}, {Dai}, \& {Khochfar}}]{shankar10a}
{Shankar}, F., {Marulli}, F., {Bernardi}, M., {et~al.} 2010, \mnras, 405, 948

\bibitem[{{Shankar} {et~al.}(2010){Shankar}, {Marulli}, {Bernardi}, {Dai},
  {Hyde}, \& {Sheth}}]{shankar10}
{Shankar}, F., {Marulli}, F., {Bernardi}, M., {et~al.} 2010, \mnras, 403, 117

\bibitem[{{Shen} {et~al.}(2003){Shen}, {Mo}, {White}, {Blanton}, {Kauffmann},
  {Voges}, {Brinkmann}, \& {Csabai}}]{shen03}
{Shen}, S., {Mo}, H.~J., {White}, S.~D.~M., {et~al.} 2003, \mnras, 343, 978

\bibitem[{{Simard} {et~al.}(2002){Simard}, {Willmer}, {Vogt}, {Sarajedini},
  {Phillips}, {Weiner}, {Koo}, {Im}, {Illingworth}, \& {Faber}}]{simard02}
{Simard}, L., {Willmer}, C.~N.~A., {Vogt}, N.~P., {et~al.} 2002, \apjs, 142, 1

\bibitem[{{Stanford} {et~al.}(1997){Stanford}, {Elston}, {Eisenhardt},
  {Spinrad}, {Stern}, \& {Dey}}]{stanford97}
{Stanford}, S.~A., {Elston}, R., {Eisenhardt}, P.~R., {et~al.} 1997, \aj, 114,
  2232

\bibitem[{{Strazzullo} {et~al.}(2010){Strazzullo}, {Rosati}, {Pannella},
  {Gobat}, {Santos}, {Nonino}, {Demarco}, {Lidman}, {Tanaka}, {Mullis},
  {Nu{\~n}ez}, {Rettura}, {Jee}, {B{\"o}hringer}, {Bender}, {Bouwens},
  {Dawson}, {Fassbender}, {Franx}, {Perlmutter}, \& {Postman}}]{strazzullo10}
{Strazzullo}, V., {Rosati}, P., {Pannella}, M., {et~al.} 2010, \aap, 524, A17

\bibitem[{{Taylor} {et~al.}(2010){Taylor}, {Franx}, {Glazebrook}, {Brinchmann},
  {van der Wel}, \& {van Dokkum}}]{taylor10}
{Taylor}, E.~N., {Franx}, M., {Glazebrook}, K., {et~al.} 2010, \apj, 720, 723

\bibitem[{{Toft} {et~al.}(2007){Toft}, {van Dokkum}, {Franx}, {Labbe},
  {F{\"o}rster Schreiber}, {Wuyts}, {Webb}, {Rudnick}, {Zirm}, {Kriek}, {van
  der Werf}, {Blakeslee}, {Illingworth}, {Rix}, {Papovich}, \&
  {Moorwood}}]{toft07}
{Toft}, S., {van Dokkum}, P., {Franx}, M., {et~al.} 2007, \apj, 671, 285

\bibitem[{{Trujillo} {et~al.}(2006){Trujillo}, {Feulner}, {Goranova}, {Hopp},
  {Longhetti}, {Saracco}, {Bender}, {Braito}, {Della Ceca}, {Drory},
  {Mannucci}, \& {Severgnini}}]{trujillo06}
{Trujillo}, I., {Feulner}, G., {Goranova}, Y., {et~al.} 2006, \mnras, 373, L36

\bibitem[{{Trujillo} {et~al.}(2007){Trujillo}, {Conselice}, {Bundy}, {Cooper},
  {Eisenhardt}, \& {Ellis}}]{trujillo07}
{Trujillo}, I., {Conselice}, C.~J., {Bundy}, K., {et~al.} 2007, \mnras, 382,
  109

\bibitem[{{Trujillo} {et~al.}(2009){Trujillo}, {Cenarro}, {de
  Lorenzo-C{\'a}ceres}, {Vazdekis}, {de la Rosa}, \& {Cava}}]{trujillo09}
{Trujillo}, I., {Cenarro}, A.~J., {de Lorenzo-C{\'a}ceres}, A., {et~al.} 2009,
  \apjl, 692, L118

\bibitem[{{Valentinuzzi} {et~al.}(2010{\natexlab{a}}){Valentinuzzi}, {Fritz},
  {Poggianti}, {Cava}, {Bettoni}, {Fasano}, {D'Onofrio}, {Couch}, {Dressler},
  {Moles}, {Moretti}, {Omizzolo}, {Kj{\ae}rgaard}, {Vanzella}, \&
  {Varela}}]{valentinuzzi10}
{Valentinuzzi}, T., {Fritz}, J., {Poggianti}, B.~M., {et~al.}
  2010{\natexlab{a}}, \apj, 712, 226

\bibitem[{{Valentinuzzi} {et~al.}(2010{\natexlab{b}}){Valentinuzzi},
  {Poggianti}, {Saglia}, {Arag{\'o}n-Salamanca}, {Simard},
  {S{\'a}nchez-Bl{\'a}zquez}, {D'onofrio}, {Cava}, {Couch}, {Fritz}, {Moretti},
  \& {Vulcani}}]{valentinuzzi10a}
{Valentinuzzi}, T., {Poggianti}, B.~M., {Saglia}, R.~P., {et~al.}
  2010{\natexlab{b}}, \apjl, 721, L19

\bibitem[{{van der Wel} {et~al.}(2008){van der Wel}, {Holden}, {Zirm}, {Franx},
  {Rettura}, {Illingworth}, \& {Ford}}]{van-der-wel08}
{van der Wel}, A., {Holden}, B.~P., {Zirm}, A.~W., {et~al.} 2008, \apj, 688, 48

\bibitem[{{van Dokkum} \& {Franx}(2001)}]{van-dokkum01a}
{van Dokkum}, P.~G. \& {Franx}, M. 2001, \apj, 553, 90

\bibitem[{{van Dokkum} {et~al.}(2008){van Dokkum}, {Franx}, {Kriek}, {Holden},
  {Illingworth}, {Magee}, {Bouwens}, {Marchesini}, {Quadri}, {Rudnick},
  {Taylor}, \& {Toft}}]{van-dokkum08}
{van Dokkum}, P.~G., {Franx}, M., {Kriek}, M., {et~al.} 2008, \apjl, 677, L5

\bibitem[{{van Dokkum} {et~al.}(2010){van Dokkum}, {Whitaker}, {Brammer},
  {Franx}, {Kriek}, {Labb{\'e}}, {Marchesini}, {Quadri}, {Bezanson},
  {Illingworth}, {Muzzin}, {Rudnick}, {Tal}, \& {Wake}}]{van-dokkum10}
{van Dokkum}, P.~G., {Whitaker}, K.~E., {Brammer}, G., {et~al.} 2010, \apj,
  709, 1018

\bibitem[{{Williams} {et~al.}(2010){Williams}, {Quadri}, {Franx}, {van Dokkum},
  {Toft}, {Kriek}, \& {Labb{\'e}}}]{williams10}
{Williams}, R.~J., {Quadri}, R.~F., {Franx}, M., {et~al.} 2010, \apj, 713, 738

\bibitem[{{Wuyts} {et~al.}(2010){Wuyts}, {Cox}, {Hayward}, {Franx},
  {Hernquist}, {Hopkins}, {Jonsson}, \& {van Dokkum}}]{wuyts10}
{Wuyts}, S., {Cox}, T.~J., {Hayward}, C.~C., {et~al.} 2010, \apj, 722, 1666

\bibitem[{{York} {et~al.}(2000){York}, {Adelman}, {Anderson}, {Anderson},
  {Annis}, {Bahcall}, {Bakken}, {Barkhouser}, {Bastian}, {Berman}, {Boroski},
  {Bracker}, {Briegel}, {Briggs}, {Brinkmann}, {Brunner}, {Burles}, {Carey},
  {Carr}, {Castander}, {Chen}, {Colestock}, {Connolly}, {Crocker}, {Csabai},
  {Czarapata}, {Davis}, {Doi}, {Dombeck}, {Eisenstein}, {Ellman}, {Elms},
  {Evans}, {Fan}, {Federwitz}, {Fiscelli}, {Friedman}, {Frieman}, {Fukugita},
  {Gillespie}, {Gunn}, {Gurbani}, {de Haas}, {Haldeman}, {Harris}, {Hayes},
  {Heckman}, {Hennessy}, {Hindsley}, {Holm}, {Holmgren}, {Huang}, {Hull},
  {Husby}, {Ichikawa}, {Ichikawa}, {Ivezi{\'c}}, {Kent}, {Kim}, {Kinney},
  {Klaene}, {Kleinman}, {Kleinman}, {Knapp}, {Korienek}, {Kron}, {Kunszt},
  {Lamb}, {Lee}, {Leger}, {Limmongkol}, {Lindenmeyer}, {Long}, {Loomis},
  {Loveday}, {Lucinio}, {Lupton}, {MacKinnon}, {Mannery}, {Mantsch}, {Margon},
  {McGehee}, {McKay}, {Meiksin}, {Merelli}, {Monet}, {Munn}, {Narayanan},
  {Nash}, {Neilsen}, {Neswold}, {Newberg}, {Nichol}, {Nicinski}, {Nonino},
  {Okada}, {Okamura}, {Ostriker}, {Owen}, {Pauls}, {Peoples}, {Peterson},
  {Petravick}, {Pier}, {Pope}, {Pordes}, {Prosapio}, {Rechenmacher}, {Quinn},
  {Richards}, {Richmond}, {Rivetta}, {Rockosi}, {Ruthmansdorfer}, {Sandford},
  {Schlegel}, {Schneider}, {Sekiguchi}, {Sergey}, {Shimasaku}, {Siegmund},
  {Smee}, {Smith}, {Snedden}, {Stone}, {Stoughton}, {Strauss}, {Stubbs},
  {SubbaRao}, {Szalay}, {Szapudi}, {Szokoly}, {Thakar}, {Tremonti}, {Tucker},
  {Uomoto}, {Vanden Berk}, {Vogeley}, {Waddell}, {Wang}, {Watanabe},
  {Weinberg}, {Yanny}, \& {Yasuda}}]{york00}
{York}, D.~G., {Adelman}, J., {Anderson}, Jr., J.~E., {et~al.} 2000, \aj, 120,
  1579

\bibitem[{{Zirm} {et~al.}(2007){Zirm}, {van der Wel}, {Franx}, {Labb{\'e}},
  {Trujillo}, {van Dokkum}, {Toft}, {Daddi}, {Rudnick}, {Rix},
  {R{\"o}ttgering}, \& {van der Werf}}]{zirm07}
{Zirm}, A.~W., {van der Wel}, A., {Franx}, M., {et~al.} 2007, \apj, 656, 66

\end{thebibliography}
\end{document}